\documentclass[lineno]{jfm}

\usepackage{amsmath,bm}
\hypersetup{
  unicode=false,
  pdfencoding=pdfdoc,
  bookmarksnumbered=true,
  bookmarksopen=true,
  pdftitle={Shear alignment and tensorial Taylor--Aris dispersion of Brownian rods in a circular tube},
  pdfauthor={Jingsen Feng and Xu Chu}
}
\graphicspath{{fig/fig1/}{fig/fig2/}{fig/fig3/}{fig/fig4/}{fig/fig5/}{fig/fig6/}{fig/fig7/}{fig/fig8/}{fig/fig9/}{fig/fig10/}}

\newcommand{\Dbar}{\bar D}
\newcommand{\Pe}{Pe}
\newcommand{\Per}{Pe_r}
\newcommand{\dd}{\,\mathrm d}
\newcommand{\avg}[1]{\left\langle #1\right\rangle}

\lefttitle{J. Feng and X. Chu}
\righttitle{Shear alignment and tensorial Taylor--Aris dispersion}

\title{Shear alignment and tensorial Taylor--Aris dispersion of Brownian rods in a circular tube}

\author{Jingsen Feng\aff{1} \and Xu Chu\aff{1}}

\affiliation{\aff{1}Department of Engineering, University of Exeter, Exeter EX4 4QF, United Kingdom}

\corresau{Xu Chu, \email{x.chu@exeter.ac.uk}}

\makeatletter
\def\jfm@simpleoddfoot{\hbox to \textwidth{\hfill{\cppagefont\thepage}}}
\def\jfm@simpleevenfoot{\hbox to \textwidth{{\cppagefont\thepage}\hfill}}
\def\oddabsfooterflag{\jfm@simpleoddfoot}
\def\evenabsfooterflag{\jfm@simpleevenfoot}
\def\pagelimitfooter{\jfm@simpleevenfoot}
\def\ps@titlepage{\leftskip\z@\let\@mkboth\@gobbletwo\vfuzz=5\p@
  \def\@oddhead{\vbox{\vspace*{-4pt}\hbox to \textwidth{\@j@urnal\hfill}}}%
  \def\@evenhead{\vbox{\vspace*{-4pt}\hbox to \textwidth{\@j@urnal\hfill}}}%
  \def\@oddfoot{\jfm@simpleoddfoot}%
  \def\@evenfoot{\jfm@simpleevenfoot}%
  \def\sectionmark##1{}%
  \def\subsectionmark##1{}%
}
\def\ps@headings{\let\@mkboth\markboth
  \def\@oddhead{\hfill{\itshape\@righttitle}\hfill}%
  \def\@evenhead{\hfill\itshape\@lefttitle\hfill}%
  \def\@oddfoot{\jfm@simpleoddfoot}%
  \def\@evenfoot{\jfm@simpleevenfoot}%
  \def\sectionmark##1{\markboth{##1}{}}%
  \def\subsectionmark##1{\markright{##1}}%
}
\makeatother
\begin{document}
\nolinenumbers
\maketitle
%\resetlinenumber[1]\runninglinenumbers

\begin{abstract}
Brownian rods disperse in pressure-driven flow through a coupling between axial shear, anisotropic translational diffusion and Jeffery--Brownian rotation. Classical tube Taylor--Aris theory treats transverse mixing as a scalar process, and existing passive-rod reductions have mainly addressed planar geometries. A circular tube adds two ingredients: the shear strength varies with radius and freely rotating rods sample a three-dimensional orientation space. We formulate a tensorial Taylor--Aris theory for dilute axisymmetric rods in Poiseuille flow by solving the local steady orientation Fokker--Planck problem and using its second moments to close a conservative axisymmetric transport equation. The long-wave reduction shows how each part of the diffusion tensor enters the one-dimensional limit. The radial diffusivity sets the invariant cross-sectional measure and the cell problem for the leading Taylor coefficient; the radial--axial component produces an inverse-P{\'e}clet correction to the migration speed; the axial component gives the direct diffusivity. The central mechanism is the streamwise alignment generated in high-shear annular layers. Alignment reduces radial diffusivity there, shifts the long-time sampling of the velocity profile toward slower streamlines, and amplifies the radial cell response. In strong shear this raises the Taylor coefficient by about \(23\%\) for aspect ratio \(p=1000\) and by about \(30\%\) in the infinitely slender limit, approaching the fully aligned bound. Direct simulations of the full tensorial equation validate the asymptotic coefficients. The same radial mixing operator also gives a Sturm--Liouville spectral model that tracks finite-time relaxation from different radial injections to the long-time Taylor regime.
\end{abstract}

\section{Introduction}

Taylor--Aris dispersion describes the longitudinal spreading produced when transverse diffusion samples a shear flow. In a circular tube, a scalar solute crosses the parabolic velocity profile by radial diffusion, and the long-time concentration evolves as a one-dimensional cloud with mean speed equal to the area-averaged flow and an enhanced axial diffusivity \citep{taylor1953,aris1956}. With the centreline velocity and tube radius used for scaling, the classical spherical-particle Taylor coefficient is \(1/192\). Subsequent work has extended this picture to finite-time spreading, shaped channels, boundary effects, pulsatile or actively controlled walls and colloidal transport in microfluidic or biological settings \citep{vedel2014,aminian2016,marbach2019,salerno2020,lee2021,alessio2022,chang2023,guan2024}. Generalized Taylor dispersion gives the corresponding cell-problem formulation for more complex transport operators and heterogeneous diffusivities in channels \citep{ramirez2006,alexandre2021}. The question considered below is how this cell problem changes when transverse mixing is set by orientational statistics.

For non-spherical Brownian particles, transverse sampling is coupled to orientation. The rotation of an ellipsoid in a linear Stokes flow follows Jeffery dynamics \citep{jeffery1922}, while Brownian rotary diffusion turns Jeffery's orbit family into a shear-dependent probability density on orientation space \citep{leal1971,hinch1972,hinch1973,brenner1974}. In simple shear, slender particles spend longer near streamwise alignment as the rotational P{\'e}clet number increases \citep{hinch1972,stover1992,leahy2015}; related orientational transport has been measured and modelled through enhanced rotational diffusion in shear \citep{leahy2013,leahy2015,peng2024} and through the dynamics of individual Brownian rods in microchannels \citep{zottl2019}. Pressure-driven flows add spatial variation to this local orientation bias and can produce lateral migration when finite-size effects, wall interactions or non-local stresses are retained \citep{nitsche1997,schiek1997}.

The translational Brownian motion of a rod is already anisotropic at zero shear: the diffusivity parallel to the particle axis differs from that in the two transverse directions, with resistance functions set by the aspect ratio \citep{tirado1979,tirado1984}. Single-particle experiments on ellipsoidal and other shaped colloids have made the coupling between particle geometry, translational diffusion and rotational diffusion directly observable \citep{han2006,han2009,chakrabarty2013,kraft2013}. A shear-biased orientation distribution therefore makes the orientation-averaged translational diffusivity a tensor in the laboratory frame. Its diagonal entries distinguish radial, azimuthal and axial spreading, and an inclined second moment in the shear plane generates a radial--axial coefficient. A single scalar transport rate hides the directional roles needed in a Taylor--Aris reduction.

\citet{dehkharghani2019} showed that orientation-controlled dispersion also appears in complex and active suspensions. Experiments and simulations of swimming cells in shear and Poiseuille flows show that orientation can suppress cross-stream motility, change drift and concentrate particles in particular shear regions \citep{zottl2012,zottl2013,rusconi2014,croze2013}. Theories for pressure-driven active suspensions, gyrotactic pipe flow and active Brownian particles have quantified the effects of particle shape, wall accumulation, upstream swimming, anisotropic diffusion and transient initial conditions on dispersivity \citep{ezhilan2015,chilukuri2015,jiang2019,jiang2020,peng2020,wang2021,jiang2021,guan2024anisodiff,wang2025}. These studies show that orientational bias can dominate confined shear transport. The active-particle setting then mixes this bias with swimming, wall accumulation and imposed drifts. The passive dilute rods considered here isolate the contribution from Jeffery--Brownian orientation statistics and anisotropic translational diffusion.

The closest precursor is the two-dimensional theory and Monte Carlo study of \citet{kumar2021}, who considered passive elongated Brownian rods in planar Poiseuille flow with one rotational degree of freedom. They showed that Jeffery alignment reduces the effective cross-stream diffusivity and enhances longitudinal Taylor dispersion relative to a sphere with the same orientationally averaged diffusivity. \citet{khair2022} subsequently derived asymptotic approximations at small and large rotational P{\'e}clet numbers, clarifying the limiting scalings of the mean speed and dispersivity. These planar passive-rod reductions leave open the corresponding circular-tube problem, where the geometry changes in two essential ways. In dimensionless form, the local shear strength varies across the cross-section as \(q=\Per r\), where \(\Per\) is the rotational P{\'e}clet number, and freely rotating rods sample the full three-dimensional orientation space. The orientation-averaged diffusion tensor then contains distinct radial, azimuthal and axial diagonal components, denoted \(D_{rr}\), \(D_{\phi\phi}\) and \(D_{zz}\), and a radial--axial cross coefficient \(D_{rz}\).

In a circular tube, the orientation-averaged diffusion tensor enters the reduction component by component. In a Taylor--Aris reduction, \(D_{rr}\) controls exchange between fast and slow streamlines and enters the leading cell problem; \(D_{zz}\) gives direct molecular spreading along the tube; and \(D_{rz}\) couples axial gradients to radial fluxes and radial gradients to axial fluxes through the conservative transport equation and the no-flux wall condition. In the high-axial-P{\'e}clet-number limit, the Taylor contribution scales quadratically with the axial P{\'e}clet number, while direct axial diffusion and cross-diffusive corrections enter at different orders.

The tube extension must therefore retain this component-wise structure rather than replacing the orientation-averaged tensor by a single scalar diffusivity. Since \(q=\Per r\), the steady orientation distribution and the effective tensor components \(D_{rr}\), \(D_{zz}\) and \(D_{rz}\) vary with radius. Variable diffusivity and anisotropic transport are known to modify invariant measures and flux balances \citep{ramirez2006,alexandre2021,guan2024anisodiff}, while shear-induced migration models show how gradients in particle stresses or diffusivities redistribute material across a channel \citep{phillips1992,nott1994,nitsche1997,schiek1997}. The dilute rod problem considered here has no phenomenological migration law. The conservative diffusive flux contains the derivative of \(D_{rr}(r)c\), so zero radial flux gives \(D_{rr}(r)c=\) constant at leading order. The cross-sectional sampling measure becomes \(rD_{rr}^{-1}(r)\,\dd r\), and the off-diagonal coefficient \(D_{rz}(r)\) enters through conservative cross-fluxes and boundary terms. What remains to be supplied is a reduction that retains this invariant measure, the radial--axial cross-diffusion and the direct axial diffusivity for Jeffery--Brownian rods in tube Poiseuille flow.

We therefore formulate such a long-wave Taylor--Aris reduction. The steady local orientation Fokker--Planck equation is solved for freely rotating axisymmetric rods at prescribed shear strength, and its second moments give \(D_{rr}^{\rm loc}\), \(D_{zz}^{\rm loc}\), \(D_{\phi\phi}^{\rm loc}\) and \(D_{rz}^{\rm loc}\). Mapping the local shear parameter to the tube radius gives the full axisymmetric tensorial transport equation for the orientation-averaged concentration. The long-wave expansion yields the leading mean migration speed, the \(D_{rz}\)-driven speed correction at inverse axial-P{\'e}clet order, the direct axial contribution from \(D_{zz}\), and the leading Taylor coefficient determined by the radial cell problem with \(D_{rr}(r)\). The same radial mixing operator is then used before the long-time limit by expanding the weighted concentration \(D_{rr}(r)c\) in its Sturm--Liouville modes. This reduced-order spectral model propagates arbitrary axisymmetric radial injections, tracks the decay of radial memory, and recovers the cell-problem Taylor coefficient in the long-time limit. Direct numerical solutions of the conservative tensorial equation validate both the asymptotic coefficients and the spectral moment predictions.

The argument below follows this sequence. Sections~\ref{sec:problem-formulation} and~\ref{sec:local-closure} define the local orientation closure and the conservative tensorial transport equation. Section~\ref{sec:taylor-aris-reduction} derives the long-time Taylor--Aris reduction, including the invariant radial measure and the separate contributions from \(D_{rr}\), \(D_{rz}\) and \(D_{zz}\). Sections~\ref{sec:long-time-coefficients} and~\ref{sec:spectral-representation} compare the asymptotic coefficients with direct simulations of the full tensorial equation and use the same radial operator to describe pre-asymptotic spreading for different radial initial distributions.

\begin{figure}
  \centering
  \includegraphics[width=0.98\linewidth]{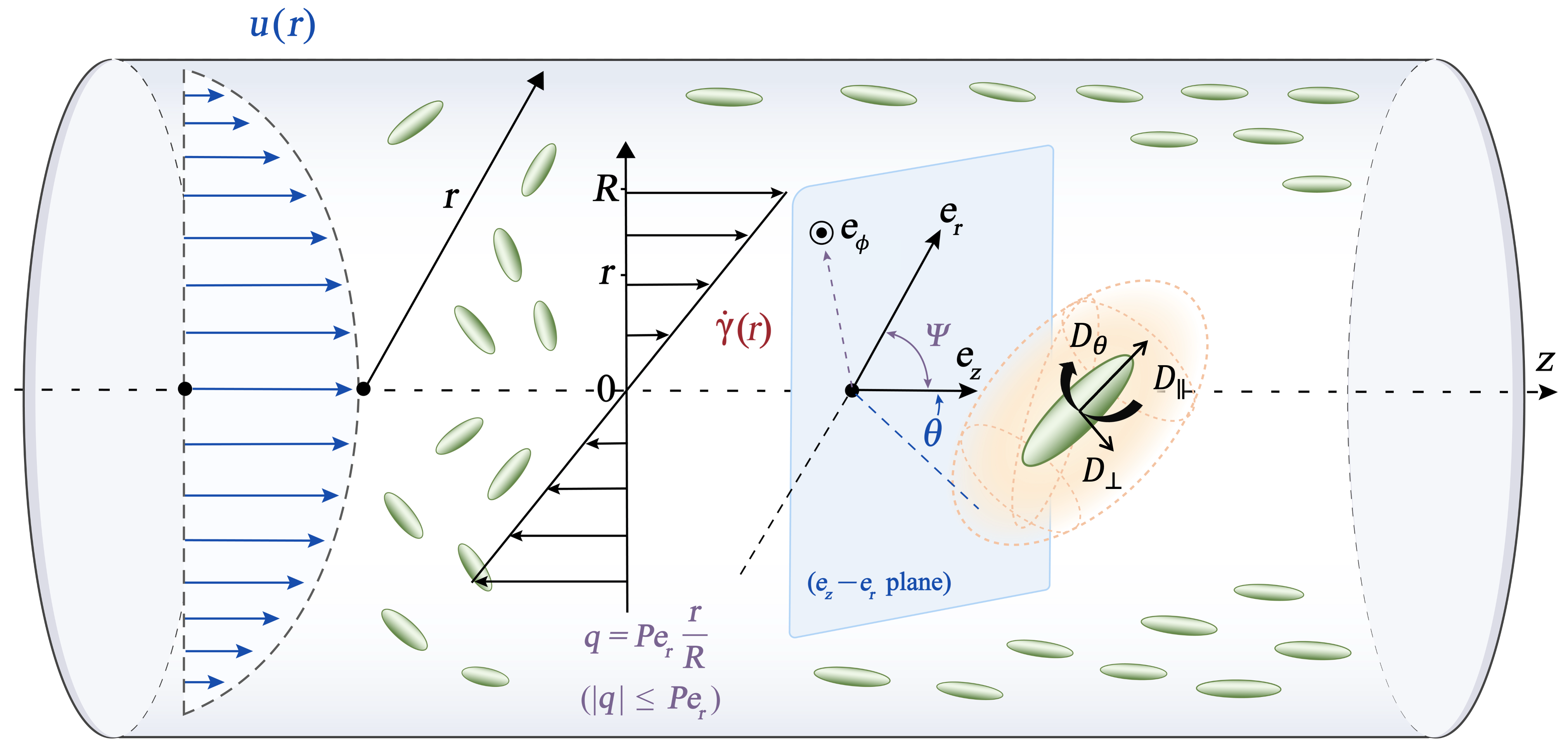}
  \caption{Geometry of freely rotating rod-like particles in tube shear flow. The left part shows the axial velocity profile \(u(r)\) and the radius-dependent shear rate \(\dot\gamma(r)\). The middle part indicates the radial position \(r\), tube radius \(R\) and local shear parameter \(q=\Per r/R\). The right inset gives the local orthonormal basis \((\hat{\bm e}_z,\hat{\bm e}_r,\hat{\bm e}_\phi)\), the shear plane and the rod-axis orientation \(\bm p\). The angle \(\theta\) measures the in-plane inclination, and \(\psi\) measures the out-of-plane inclination. In the schematic \(R\) is the dimensional tube radius, denoted by \(a\) in the text; after non-dimensionalization \(r/R\) is written as \(r\), so that \(q=\Per r\).}
  \label{fig:schematic}
\end{figure}

\section{Problem formulation and non-dimensionalization}
\label{sec:problem-formulation}

Consider a dilute suspension of rod-like particles in an infinitely long circular tube. The axial coordinate is \(z\), the dimensional radial coordinate is \(r^\ast\), and the tube radius is \(a\). The dimensional Poiseuille velocity field is
\begin{equation}
  \bm u_f^\ast(r^\ast)=U\left(1-\frac{r^{\ast 2}}{a^2}\right)\hat{\bm e}_z ,
  \qquad 0\le r^\ast\le a ,
\label{eq:dimensional-flow}
\end{equation}
where \(U\) is the centreline speed. The local shear rate is
\begin{equation}
  \dot\gamma^\ast(r^\ast)=\frac{\dd}{\dd r^\ast}
  \left[
  U\left(1-\frac{r^{\ast 2}}{a^2}\right)
  \right]
  =-\frac{2Ur^\ast}{a^2}.
\label{eq:shear-rate-dimensional}
\end{equation}
In the local orthonormal basis \((\hat{\bm e}_z,\hat{\bm e}_r,\hat{\bm e}_\phi)\), the velocity gradient can be written as
\begin{equation}
  \nabla\bm u_f^\ast
  =
  \dot\gamma^\ast\,\hat{\bm e}_z\hat{\bm e}_r,
\end{equation}
with symmetric and antisymmetric parts
\begin{equation}
  \bm E^\ast=
  \frac{\dot\gamma^\ast}{2}
  \left(\hat{\bm e}_z\hat{\bm e}_r+\hat{\bm e}_r\hat{\bm e}_z\right),
  \qquad
  \bm W^\ast=
  \frac{\dot\gamma^\ast}{2}
  \left(\hat{\bm e}_z\hat{\bm e}_r-\hat{\bm e}_r\hat{\bm e}_z\right).
\end{equation}
Here \(\bm E^\ast\) and \(\bm W^\ast\) are the symmetric and antisymmetric
parts of the velocity gradient, respectively. The tensor \(\bm E^\ast\) is the
rate-of-strain tensor and describes the local deformation of a fluid element,
whereas \(\bm W^\ast\) is the spin tensor and describes its local rigid-body
rotation, corresponding to one half of the local vorticity.
The closure is formulated for a dilute, passive suspension, so that interparticle interactions and particle feedback on the Poiseuille flow are neglected. At the macroscopic level, the concentration field is axisymmetric. At the particle level, each rod axis rotates freely in the three-dimensional orientation space according to Jeffery dynamics, supplemented by Brownian rotational diffusion at low Reynolds number \citep{jeffery1922,leal1971,hinch1972,brenner1974}; the \(\bm E^\ast\) contribution gives the shape-dependent strain response responsible for shear-induced alignment, whereas the \(\bm W^\ast\) contribution rotates the rod with the local fluid spin. Wall-induced hydrodynamic corrections, finite-size exclusion and lubrication effects are omitted.

The translational diffusivity along the rod axis is denoted by \(D_\parallel\), and that in any direction perpendicular to the rod axis by \(D_\perp\). The diffusion tensor is normalized by the three-dimensional mean translational diffusivity
\begin{equation}
  \Dbar=\frac{D_\parallel+2D_\perp}{3}.
\label{eq:Dbar}
\end{equation}
For the numerical examples, the ratios \(D_\parallel/\Dbar\), \(D_\perp/\Dbar\) and the rotational diffusivity \(D_\theta\) are set from the Perrin resistance functions for a prolate spheroid \citep{perrin1934,perrin1936}. For \(p>1\), suppressing the dimensional Perrin prefactors, the shape-dependent parts are
\begin{subequations}
\label{eq:perrin-trans}
\begin{align}
  D_\parallel
  &=
  p\left[
  -\frac{2p}{p^2-1}
  +\frac{2p^2-1}{(p^2-1)^{3/2}}
  \log\!\left(\frac{p+\sqrt{p^2-1}}{p-\sqrt{p^2-1}}\right)
  \right],\\
  D_\perp
  &=
  p\left[
  \frac{p}{p^2-1}
  +\frac{2p^2-3}{(p^2-1)^{3/2}}
  \log\!\left(p+\sqrt{p^2-1}\right)
  \right],\\
  D_\theta
  &=
  \frac{p^4}{p^4-1}
  \left[
  \frac{2p^2-1}{p\sqrt{p^2-1}}
  \log\!\left(p+\sqrt{p^2-1}\right)-1
  \right],
\end{align}
\end{subequations}
Together with \eqref{eq:Dbar}, these expressions give the dimensionless
translational ratios used below. The spherical limit has
\(D_\parallel/\Dbar=D_\perp/\Dbar=1\), while the infinitely slender limit gives
\(D_\parallel/\Dbar\to1.5\) and \(D_\perp/\Dbar\to0.75\). The continuum
closure uses these dimensionless diffusivity ratios as inputs; alternative
microscopic resistance models can be incorporated by replacing the ratios.

Lengths are scaled with \(a\), and time is scaled with \(a^2/\Dbar\). After dropping hats on dimensionless variables, the velocity profile is
\begin{equation}
  u(r)=1-r^2,\qquad 0\le r\le1,
\label{eq:poiseuille-nondim}
\end{equation}
and the axial translational P{\'e}clet number and rotational P{\'e}clet number are defined by
\begin{equation}
  \Pe=\frac{Ua}{\Dbar},\qquad
  \Per=\frac{U}{aD_\theta},
\label{eq:peclet-definitions}
\end{equation}
where \(D_\theta\) is the rotational diffusivity. The signed shear parameter and its magnitude are
\begin{equation}
  \chi(r)=\frac{\dot\gamma^\ast(r^\ast)}{2D_\theta}=-\Per r,
  \qquad
  q(r)=|\chi(r)|=\Per r .
\label{eq:local-q}
\end{equation}
The magnitude \(q\) measures the strength of the local Jeffery drift relative
to Brownian rotational diffusion, and \(\chi\) carries the shear sign. The
orientational dynamics are assumed to relax rapidly relative to radial
transport. With \(t_\theta\sim D_\theta^{-1}\) and \(t_r\sim a^2/\Dbar\), this
scale separation requires
\begin{equation}
  \epsilon_\theta=\frac{t_\theta}{t_r}
  =\frac{\Dbar}{a^2D_\theta}
  =\frac{\Per}{\Pe}
  \ll1 .
\label{eq:orientation-separation}
\end{equation}
Under this scale separation, the spatial transport equation is closed by the
steady orientation distribution corresponding to the local shear strength. The
local orientation solutions are tabulated as functions of \(q\); the sign of
the tube shear is restored when the radial--axial coefficient \(D_{rz}\) is
mapped into the tube.

The particle aspect ratio is denoted by \(p\), with shape parameter
\begin{equation}
  \beta=\frac{p^2-1}{p^2+1}.
\label{eq:beta}
\end{equation}
Here \(\beta=0\) corresponds to a sphere, and \(\beta\to1\) to an infinitely slender rod. The rod-axis unit vector \(\bm p\) is parameterized in the local cylindrical basis as
\begin{equation}
  \bm p
  =
  \cos\theta\cos\psi\,\hat{\bm e}_z
  +\sin\theta\cos\psi\,\hat{\bm e}_r
  +\sin\psi\,\hat{\bm e}_\phi .
\label{eq:p-param}
\end{equation}
Thus
\begin{equation}
  p_z=\cos\theta\cos\psi,\qquad
  p_r=\sin\theta\cos\psi,\qquad
  p_\phi=\sin\psi,
\end{equation}
and the solid-angle element is
\begin{equation}
  \dd\Omega=\cos\psi\,\dd\psi\,\dd\theta .
\end{equation}
With \(\mu=\sin\psi\), \(\dd\Omega=\dd\mu\,\dd\theta\), and the angular domain is \(0\le\theta<2\pi\), \(-1\le\mu\le1\). Since the rod axis satisfies \(\bm p\equiv-\bm p\), the orientation distribution has the corresponding head-tail symmetry.

Figure~\ref{fig:schematic} summarizes the geometry and local orientation variables. The Poiseuille velocity profile gives a radially varying local shear rate, with strength \(q=\Per r\) after non-dimensionalization. The rod axis \(\bm p\) rotates freely in three-dimensional orientation space; \(\theta\) measures its inclination within the \(\hat{\bm e}_z\)-\(\hat{\bm e}_r\) shear plane, and \(\psi\) measures the out-of-plane inclination. Shear-induced streamwise alignment changes the spatial contributions of diffusion along and transverse to the long axis, producing a radius-dependent anisotropic diffusion tensor.

\section{Local orientation closure and axisymmetric transport equation}
\label{sec:local-closure}

A steady orientational distribution \(g_q\) is obtained at each radius from the local shear. The shear parameter \(q\) fixes this distribution, and its second-order moments determine the effective local diffusion tensor. The \(r\)-\(z\) block of that tensor enters the axisymmetric transport equation for the orientation-averaged concentration \(c(r,z,t)\), while the azimuthal component \(D_{\phi\phi}\) is retained as a diagnostic of the three-dimensional orientational statistics and as a trace-identity check. The resulting local closure provides the coefficients \(D_{rr}^{\rm loc}\), \(D_{zz}^{\rm loc}\), \(D_{\phi\phi}^{\rm loc}\) and \(D_{rz}^{\rm loc}\).

\subsection{\texorpdfstring{Orientation Fokker--Planck equation}{Orientation Fokker-Planck equation}}

Within a small neighbourhood of a fixed tube radius, the local velocity gradient is that of a simple shear. The rod axis satisfies Jeffery's equation \citep{jeffery1922}
\begin{equation}
  \dot{\bm p}
  =
  \bm W^\ast\cdot\bm p
  +\beta\left[
  \bm E^\ast\cdot\bm p
  -(\bm p\cdot\bm E^\ast\cdot\bm p)\bm p
  \right].
\label{eq:jeffery-vector}
\end{equation}
With the cylindrical basis and shear-rate convention in \eqref{eq:shear-rate-dimensional}, projection of \eqref{eq:jeffery-vector} onto \((\theta,\psi)\) gives
\begin{equation}
  \dot\theta=-\dot\gamma^\ast \Omega_\theta,\qquad
  \dot\psi=-\dot\gamma^\ast \Omega_\psi,
\end{equation}
where
\begin{equation}
  \Omega_\theta=\frac12(1-\beta\cos 2\theta),
  \qquad
  \Omega_\psi=\frac14\beta\sin2\theta\sin2\psi .
\label{eq:omega-theta-psi}
\end{equation}
For \(\beta=0\), Jeffery's equation reduces to rigid rotation by the vorticity and the steady orientational distribution remains uniform. As \(\beta\) increases, the magnitude of the angular drift decreases near streamwise alignment. The steady probability density therefore increases near these orientations, with rotational diffusion setting the finite width of the peak. With \(\mu=\sin\psi\),
\begin{equation}
  \dot\mu=-\dot\gamma^\ast\Omega_\mu,\qquad
  \Omega_\mu=\frac12\beta \mu(1-\mu^2)\sin2\theta .
\end{equation}

Let \(g(\theta,\mu;q)\) be the steady orientation probability density for a reference simple shear with \(\dot\gamma^\ast/(2D_\theta)=q>0\). The tube shear has the opposite sign, as recorded by \(\chi=-q\) in \eqref{eq:local-q}; this sign is applied to the off-diagonal transport coefficient in \eqref{eq:radial-mapping}. In \((\theta,\mu)\) coordinates, the steady Fokker--Planck equation for the reference branch is \citep{leal1971,hinch1972,brenner1974}
\begin{equation}
\begin{split}
  &\frac{1}{1-\mu^2}\frac{\partial^2 g}{\partial\theta^2}
  +\frac{\partial}{\partial\mu}
  \left[(1-\mu^2)\frac{\partial g}{\partial\mu}\right]\\
  &\quad
  +2q\left[
  \frac{\partial}{\partial\theta}(\Omega_\theta g)
  +\frac{\partial}{\partial\mu}(\Omega_\mu g)
  \right]=0,
\end{split}
\label{eq:orientation-fp}
\end{equation}
with periodicity in \(\theta\), regularity at \(\mu=\pm1\), normalization
\begin{equation}
  \int_0^{2\pi}\int_{-1}^{1}
  g(\theta,\mu;q)\,\dd\mu\,\dd\theta=1,
\label{eq:g-normalization}
\end{equation}
and fore-aft symmetry
\begin{equation}
  g(\theta+\pi,-\mu;q)=g(\theta,\mu;q).
\label{eq:g-head-tail}
\end{equation}
Equation~\eqref{eq:orientation-fp}, together with these conditions, defines the local orientation closure. In the tube calculation the tabulated solutions are sampled at \(q=\Per r\).

\subsection{Effective diffusion tensor}

The orientation distribution enters the spatial transport equation through the translational diffusion tensor. For a fixed orientation, the anisotropic part of the translational diffusion tensor is proportional to \(\bm p\bm p\). Hence the local closure depends only on the second-order moments \(\avg{p_i p_j}_q\).
For a single orientation \(\bm p\), the dimensionless translational diffusion tensor is \citep{perrin1934,perrin1936}
\begin{equation}
  \bm D(\bm p)
  =
  \frac{D_\perp}{\Dbar}\bm I
  +\frac{D_\parallel-D_\perp}{\Dbar}\bm p\bm p .
\label{eq:single-orientation-diffusion}
\end{equation}
Averaging over the local orientation distribution is denoted by
\begin{equation}
  \avg{f}_q=
  \int_0^{2\pi}\int_{-1}^{1}
  f(\theta,\mu)g(\theta,\mu;q)\,\dd\mu\,\dd\theta .
\label{eq:local-average}
\end{equation}
The local tensor components are
\begin{subequations}
\label{eq:local-tensor-definitions}
\begin{align}
  D_{rr}^{\rm loc}(q)&=
  \frac{D_\perp}{\Dbar}
  +\frac{D_\parallel-D_\perp}{\Dbar}\avg{p_r^2}_q,
  \label{eq:local-tensor-Drr}\\
  D_{zz}^{\rm loc}(q)&=
  \frac{D_\perp}{\Dbar}
  +\frac{D_\parallel-D_\perp}{\Dbar}\avg{p_z^2}_q,
  \label{eq:local-tensor-Dzz}\\
  D_{\phi\phi}^{\rm loc}(q)&=
  \frac{D_\perp}{\Dbar}
  +\frac{D_\parallel-D_\perp}{\Dbar}\avg{p_\phi^2}_q,
  \label{eq:local-tensor-Dphiphi}\\
  D_{rz}^{\rm loc}(q)&=
  \frac{D_\parallel-D_\perp}{\Dbar}\avg{p_rp_z}_q .
  \label{eq:local-tensor-Drz}
\end{align}
\end{subequations}
Here \(D_{rr}^{\rm loc}\) is the radial mixing rate controlling cross-stream sampling, \(D_{zz}^{\rm loc}\) is the direct axial diffusivity, and \(D_{rz}^{\rm loc}\) arises from the inclined moment \(\avg{p_rp_z}_q\) in the shear plane. The component \(D_{\phi\phi}^{\rm loc}\) characterizes the out-of-plane orientation statistics and provides a trace-identity check.
Since \(p_r^2+p_z^2+p_\phi^2=1\), the components satisfy the trace identity
\begin{equation}
  D_{rr}^{\rm loc}+D_{zz}^{\rm loc}+D_{\phi\phi}^{\rm loc}=3.
\label{eq:trace-identity}
\end{equation}
The \(r\)-\(z\) block of the effective tensor must also remain positive definite:
\begin{equation}
  D_{rr}^{\rm loc}>0,\qquad
  D_{rr}^{\rm loc}D_{zz}^{\rm loc}-(D_{rz}^{\rm loc})^2>0.
\label{eq:positive-definite}
\end{equation}
These relations constrain the closure coefficients and provide consistency checks for the subsequent tube transport equation.

\begin{figure}
  \centering
  \includegraphics[width=0.92\linewidth]{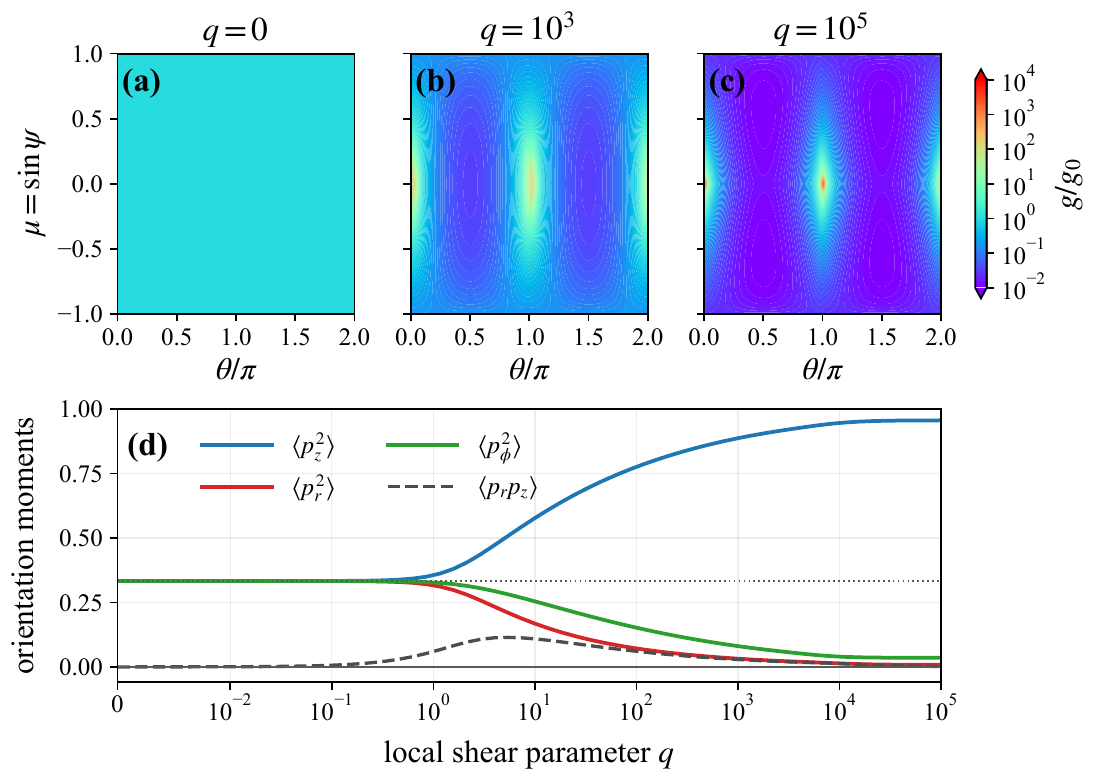}
  \caption{Three-dimensional orientational bias induced by local shear. The parameters are \(p=1000\), corresponding to \(\beta=0.999998\), \(D_\parallel/\Dbar=1.4013\) and \(D_\perp/\Dbar=0.7993\). Panel (a) shows the probability density \(g/g_0\) in \((\theta,\mu)\) coordinates at \(q=0\). Panel (b) shows the same density at \(q=10^3\). Panel (c) shows the same density at \(q=10^5\). Here \(\mu=\sin\psi\) and \(g_0=1/(4\pi)\). Panel (d) shows the second-order orientation moments \(\avg{p_z^2}\), \(\avg{p_r^2}\), \(\avg{p_\phi^2}\) and \(\avg{p_rp_z}\) as functions of \(q\).}
  \label{fig:orientation}
\end{figure}

Figure~\ref{fig:orientation} shows the effect of increasing \(q\) on the local orientational statistics. At \(q=0\), the distribution is isotropic, with \(\avg{p_r^2}=\avg{p_z^2}=\avg{p_\phi^2}=1/3\) and \(\avg{p_rp_z}=0\). Larger \(q\) shifts probability density towards streamwise alignment, increasing \(\avg{p_z^2}\) and decreasing \(\avg{p_r^2}\). For \(p=1000\), \(\avg{p_z^2}\) rises from \(0.577\) at \(q=10\) to \(0.887\) at \(q=10^3\), while \(\avg{p_r^2}\) falls from \(0.168\) to \(0.032\). The inclined moment \(\avg{p_rp_z}\) is largest at intermediate shear and decreases once the distribution becomes strongly concentrated near the axial direction.

\begin{figure}
  \centering
  \includegraphics[width=0.92\linewidth]{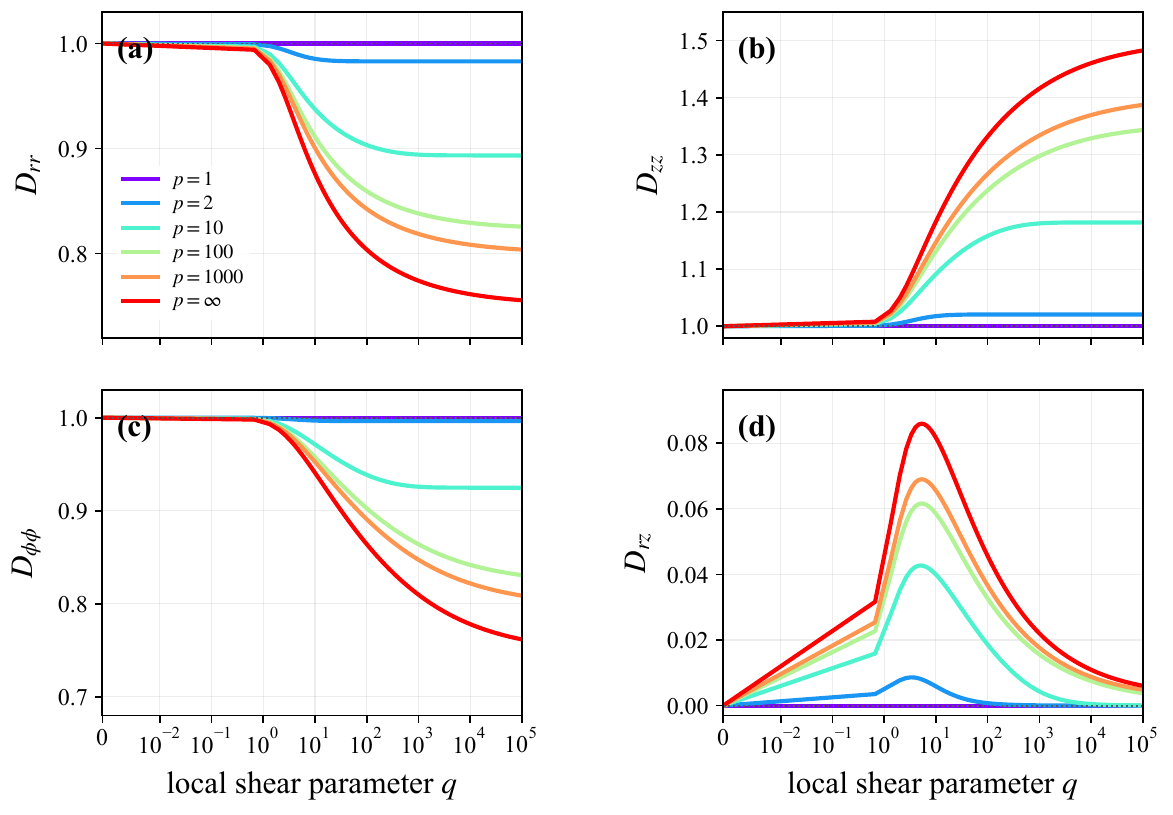}
  \caption{Variation of the local orientation-averaged transport tensor with shear parameter \(q\). Panel (a) shows \(D_{rr}^{\rm loc}\). Panel (b) shows \(D_{zz}^{\rm loc}\). Panel (c) shows \(D_{\phi\phi}^{\rm loc}\). Panel (d) shows \(D_{rz}^{\rm loc}\). The superscript ``loc'' is omitted in the panel labels. Different curves correspond to different aspect ratios \(p\). The spherical limit \(p=1\) gives \(D_{rr}=D_{zz}=D_{\phi\phi}=1\) and \(D_{rz}=0\).}
  \label{fig:local-tensor}
\end{figure}

Figure~\ref{fig:local-tensor} shows the corresponding orientation-averaged diffusivities. Spherical particles have \(D_\parallel=D_\perp\), so the translational diffusion tensor is independent of orientation. For rods, streamwise alignment reduces \(D_{rr}^{\rm loc}\) and increases \(D_{zz}^{\rm loc}\). For \(p=1000\), \(D_{rr}^{\rm loc}\) decreases by about \(20\%\) at large \(q\), whereas the peak value of \(|D_{rz}^{\rm loc}/D_{rr}^{\rm loc}|\) is \(O(10^{-1})\). After the tube mapping, \(D_{rr}(r)\) controls the first change in radial sampling and \(D_{rz}(r)\) enters as a cross-flux correction. Reversing the shear sign leaves the diagonal components unchanged and reverses \(D_{rz}^{\rm loc}\).

\subsection{Closed axisymmetric tensorial transport equation}

Under the scale separation in \eqref{eq:orientation-separation}, particles at a given radius sample the same orientational statistics as in a homogeneous simple shear with strength \(q(r)=\Per r\). Mapping this local closure back to the tube coordinates requires the sign of the Poiseuille shear. Since \(s_\gamma=\operatorname{sgn}(\dot\gamma^\ast)=-1\) for \(0<r\le1\), the diagonal components are unchanged and the radial--axial coefficient changes sign:
\begin{subequations}
\label{eq:radial-mapping}
\begin{align}
  D_{rr}(r)&=D_{rr}^{\rm loc}(q(r)),&
  D_{zz}(r)&=D_{zz}^{\rm loc}(q(r)),&
  D_{\phi\phi}(r)&=D_{\phi\phi}^{\rm loc}(q(r)),\\
  D_{rz}(r)&=s_\gamma D_{rz}^{\rm loc}(q(r)).
\end{align}
\end{subequations}
At \(r=0\), \(D_{rz}=0\) by symmetry. The component \(D_{\phi\phi}\) does not enter the axisymmetric \(r\)-\(z\) equation because there is no azimuthal concentration gradient; it is retained in the local closure diagnostics.
To emphasize the flux structure, define
\begin{equation}
  D(r)=D_{rr}(r),\qquad A(r)=D_{rz}(r),\qquad B(r)=D_{zz}(r).
\end{equation}
Let \(c(r,z,t)\) be the number concentration after local orientation averaging, and let \(n(r,z,\bm p,t)\) be the orientation-resolved number density. The conservative fluxes follow from the local-equilibrium ansatz
\begin{equation}
  n(r,z,\bm p,t)=c(r,z,t)g_{q(r)}(\bm p)
\label{eq:local-equilibrium-ansatz}
\end{equation}
inserted into the translational Smoluchowski flux and integrated over orientation \citep{brenner1974,kumar2021,khair2022}:
\begin{equation}
  J_i=\Pe u_i c-
  \int D_{ij}(\bm p)\,\partial_j
  \left[c(r,z,t)g_{q(r)}(\bm p)\right]\,\dd\Omega,
  \qquad i,j\in\{r,z\},
\label{eq:averaged-flux-derivation}
\end{equation}
where \(u_z=u(r)\) and \(u_r=0\). Since \(g_{q(r)}\) varies with radius through \(q=\Per r\), the radial derivative acts on both \(c\) and the local orientational equilibrium. This gives
\begin{align}
  J_r&=
  -\frac{\partial}{\partial r}\left[D(r)c\right]-A(r)c_z,\\
  J_z&=
  \Pe u(r)c-\frac{\partial}{\partial r}\left[A(r)c\right]-B(r)c_z \notag\\
  &=\Pe u(r)c-A(r)c_r-A'(r)c-B(r)c_z .
\label{eq:closed-fluxes}
\end{align}
The axisymmetric transport equation is therefore
\begin{equation}
  c_t+\frac1r\frac{\partial}{\partial r}(rJ_r)
  +\frac{\partial J_z}{\partial z}=0.
\label{eq:closed-pde-conservation}
\end{equation}
The radial flux may also be expanded as
\begin{equation}
  J_r=v_D(r)c-D(r)c_r-A(r)c_z,
\end{equation}
where
\begin{equation}
  v_D(r)=-\frac{\dd D}{\dd r}
\label{eq:radial-drift}
\end{equation}
is the statistical drift associated with radially inhomogeneous orientational statistics. The fluid velocity has no radial component; the radial contribution in \eqref{eq:radial-drift} comes from shear-induced variation of \(D_{rr}(r)\) within the conservative diffusive flux. In the absence of an axial gradient, \(J_r=-(Dc)_r\), and the zero-flux radial equilibrium gives \(D(r)c=\text{constant}\). This equilibrium later appears as the invariant radial measure that weights cross-sectional sampling in the Taylor--Aris reduction.

The radial boundary conditions are regularity at the axis and no penetration at the wall:
\begin{equation}
  rJ_r\to0\quad(r\to0),\qquad
  J_r(1,z,t)=0.
\label{eq:closed-boundary}
\end{equation}
Since \(J_r=-(Dc)_r-Ac_z\), the no-flux wall condition induces a compensating radial gradient when an axial concentration gradient is present. This constraint is the boundary source by which cross-diffusion enters the next-order correction to the mean migration speed.
For the asymptotic transport problem the axial coordinate is unbounded. The total cross-sectional mass, mean axial position and axial variance are computed using the axisymmetric weight \(r\):
\begin{align}
  M(t)&=2\int_0^1\int_{-\infty}^{\infty} c(r,z,t)\,r\,\dd z\,\dd r,\\
  \avg{z}(t)&=
  \frac{2}{M(t)}\int_0^1\int_{-\infty}^{\infty} z\,c(r,z,t)\,r\,\dd z\,\dd r,\\
  \sigma_z^2(t)&=
  \frac{2}{M(t)}\int_0^1\int_{-\infty}^{\infty}
  [z-\avg{z}(t)]^2c(r,z,t)\,r\,\dd z\,\dd r.
\label{eq:pde-moments}
\end{align}
For direct numerical validation, the same equation is solved on a periodic interval \(0\le z<L_z\). The interval is chosen so that the concentration packet does not interact with its periodic images during the fitting window. The mean position and variance are then evaluated using an unfolded coordinate, or an equivalent periodic copy window.

Equations~\eqref{eq:closed-fluxes} and \eqref{eq:closed-pde-conservation} are the closed two-dimensional equation solved in the direct simulations. The axial advection term \(\Pe u(r)c\) supplies the velocity contrast responsible for Taylor dispersion. The term \(-(Dc)_r\) determines radial equilibrium and the leading cell problem. The terms \(-Ac_z\) and \(-(Ac)_r=-Ac_r-A'c\) provide conservative cross-fluxes, and \(-Bc_z\) gives direct axial diffusion. The leading Taylor coefficient scaled by \(\Pe^2\) is controlled by transverse mixing, while \(Bc_{zz}\) contributes an \(O(1)\) direct diffusion to the unscaled variance.

\begin{figure}
  \centering
  \includegraphics[width=0.98\linewidth]{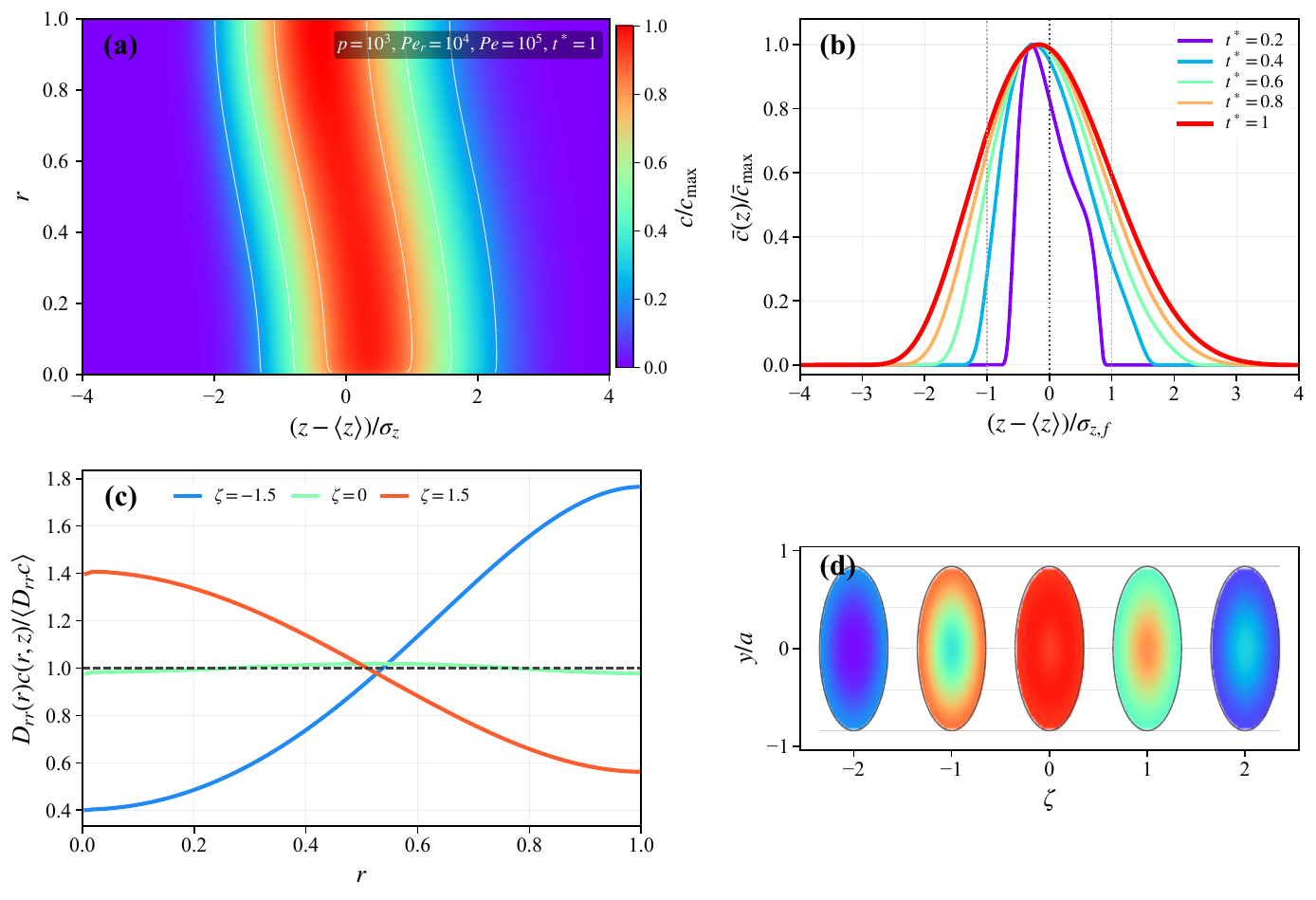}
  \caption{Representative concentration field from the closed axisymmetric tensorial transport equation. The parameters are \(p=1000\), \(\Per=10^4\) and \(\Pe=10^5\). The normalized moving coordinate is \(\zeta=(z-\langle z\rangle)/\sigma_{z,f}\), where \(\sigma_{z,f}\) is the final-time axial standard deviation. Panel (a) shows the concentration field in the moving coordinate, \(c(r,z-\langle z\rangle)/c_{\max}\). Panel (b) shows the cross-sectionally averaged axial concentration \(\bar c(z,t)=2\int_0^1 rc(r,z,t)\,\dd r\), with the horizontal coordinate normalized by \(\sigma_{z,f}\). Panel (c) gives the radial-equilibrium diagnostic \(D_{rr}(r)c(r,z)\) at selected \(\zeta\). Panel (d) shows the same axisymmetric concentration field after rotation about the tube axis, with cross-sections placed at their \(\zeta\) locations.}
  \label{fig:closed-field}
\end{figure}

The closed equation produces a concentration packet that broadens axially in the moving coordinate while retaining a distinct radial structure (Figure~\ref{fig:closed-field}). The cross-sectionally averaged profiles in Figure~\ref{fig:closed-field}(b), normalized by the same final standard deviation, reveal the growth of the packet width with time. The radial-equilibrium diagnostic in Figure~\ref{fig:closed-field}(c) tests the long-time relation \(D_{rr}c\simeq\text{constant}\), which follows from the leading-order no-flux condition; the central part of the packet is closer to equilibrium, while the tails away from the peak retain stronger radial corrections. The rotated rendering in Figure~\ref{fig:closed-field}(d) aids interpretation of the same axisymmetric field over the tube cross-section.

This field motivates the long-time reduction in the next section. Its evolution indicates which radial structures must survive the asymptotic limit: radial diffusion sets the invariant measure, axial shear supplies the velocity deviation, the variable-coefficient drift shapes the radial equilibrium, and cross-diffusion modifies the next radial correction. The effective one-dimensional coefficients must therefore be obtained from radial cell problems involving \(D(r)\) and \(A(r)\); a scalar cross-sectional average would miss the invariant measure and the conservative cross-flux.

\section{\texorpdfstring{Taylor--Aris long-time reduction}{Taylor-Aris long-time reduction}}
\label{sec:taylor-aris-reduction}

\subsection{Radial equilibrium}

The closed tensorial equation remains two-dimensional, and the long-time axial spreading is most transparent after a Taylor--Aris reduction \citep{taylor1953,aris1956,ramirez2006,alexandre2021} that preserves the radial tensor profiles generated by orientation. We consider a long-wave limit in which \(\Pe\to\infty\) while \(p\) and \(\Per\) are fixed parameters of the local orientational closure. During the axial reduction the tensor components are prescribed functions of radius:
\begin{equation}
  D(r)=D_{rr}(r),\qquad A(r)=D_{rz}(r),\qquad B(r)=D_{zz}(r).
\label{eq:D-A-def}
\end{equation}
In this section \(D(r)\) denotes the scalar radial component \(D_{rr}(r)\). The axial component \(B(r)\) contributes to the \(O(1)\) direct axial diffusivity and is kept separate from the \(\Pe^2\)-scaled Taylor coefficient. The small parameter is
\begin{equation}
  \Pe\gg1,\qquad
  \epsilon=\frac{\Per}{\Pe}=\frac{\Dbar}{a^2D_\theta}\ll1.
\end{equation}
This ordering separates orientational relaxation, radial equilibration and axial Taylor spreading. The coefficient \(\kappa(\Per,p)\) below is defined in this ordering; parameter scans in \(\Per\) are interpreted under the same condition \(\Per/\Pe\ll1\).

To follow the slow axial spreading relative to the unknown mean drift, we use the moving axial coordinate
\begin{equation}
  X=z-\Pe u_m t,
\end{equation}
and define the slow variables
\begin{equation}
  \xi=\epsilon^2X,\qquad T=\epsilon^2t,\qquad
  \epsilon=\frac{\Per}{\Pe}\ll1.
\label{eq:slow-variables}
\end{equation}
The factor \(\epsilon^2\) in the axial slow variable is chosen so that
\(\Pe\partial_z=\epsilon\Per\partial_\xi\); the shear-induced radial correction
then enters at \(O(\epsilon)\), while the reduced diffusion equation transforms
back to the \(O(\Pe^2)\) Taylor term in \eqref{eq:effective1d}.
The concentration and mean speed are expanded as
\begin{equation}
  C=C^{(0)}+\epsilon C^{(1)}+\epsilon^2C^{(2)}+\cdots,
  \qquad
  u_m=u_m^{(0)}+\epsilon u_m^{(1)}+\cdots .
\label{eq:taylor-expansions}
\end{equation}
Here \(C(r,\xi,T)\) denotes the concentration in the moving coordinate. With the diffusivity-induced drift \(v_D=-D'(r)\),
\begin{equation}
  v_D C-D C_r=-(DC)_r,
\end{equation}
the leading radial operator is
\begin{equation}
  \mathcal L_0 C
  =
  \frac1r\frac{\partial}{\partial r}
  \left[
  r\frac{\partial}{\partial r}(D C)
  \right].
\label{eq:radial-operator-C}
\end{equation}

The leading-order equation is
\begin{equation}
  \mathcal L_0 C^{(0)}=0.
\end{equation}
Multiplying by \(r\), integrating from \(0\) to \(r\), and using regularity at the axis gives
\begin{equation}
  \frac{\partial}{\partial r}\left[D(r)C^{(0)}(r,\xi,T)\right]=0.
\end{equation}
The wall no-flux condition gives the same conclusion, and hence
\begin{equation}
  C^{(0)}(r,\xi,T)=\frac{C_m(\xi,T)}{D(r)}
  .
\label{eq:C0}
\end{equation}
\(C_m\) is the radial-equilibrium amplitude. The normalization integral is
\begin{equation}
  I_0=\int_0^1 \frac{r}{D(r)}\,\dd r .
\label{eq:I0}
\end{equation}
The leading-order cross-sectional mass density is
\begin{equation}
  M(\xi,T)=2\int_0^1 rC^{(0)}(r,\xi,T)\,\dd r=2I_0C_m(\xi,T).
\end{equation}
Since \(I_0\) is independent of \(\xi\) and \(T\), the amplitude \(C_m\) and the cross-sectional mass density \(M\) obey the same reduced one-dimensional transport equation. We use \(C_m\) as the reduced concentration variable. The equilibrium relation in \eqref{eq:C0} has a direct sampling interpretation: particles spend more time in radial layers where the local radial diffusivity is smaller. The invariant cross-sectional measure is therefore the weighted measure \(rD^{-1}(r)\,\dd r\):
\begin{equation}
  \pi(r)\,\dd r=
  \frac{rD^{-1}(r)\,\dd r}
       {\displaystyle\int_0^1 sD^{-1}(s)\,\dd s}.
\label{eq:pi-measure}
\end{equation}
The classical spherical problem samples the Poiseuille velocity profile with the area measure \(r\,\dd r\). For rod-like particles, local orientational alignment changes this to \(rD^{-1}(r)\,\dd r\), consistent with generalized Taylor--Aris formulations for heterogeneous transverse transport \citep{ramirez2006,alexandre2021}, so low-\(D_{rr}\) regions gain weight in the long-time radial equilibrium. This reweighting is the first mechanism through which local alignment affects the mean speed and the Taylor coefficient.

\subsection{Mean speed and cell problem}

The next-order equation can be written as
\begin{equation}
  \mathcal L_0 C^{(1)}
  =
  \Per\,
  \frac{u(r)-u_m^{(0)}}{D(r)}
  \frac{\partial C_m}{\partial \xi}.
\label{eq:C1-equation}
\end{equation}
The solvability condition for \eqref{eq:C1-equation} follows from the null mode of the radial operator. Multiplying by \(r\) and integrating over \(0<r<1\), the left-hand side reduces to a boundary flux; regularity at the axis and no flux at the wall make this boundary contribution vanish. Hence
\begin{equation}
  \int_0^1 r\,\frac{u(r)-u_m^{(0)}}{D(r)}\,\dd r=0,
\end{equation}
and
\begin{equation}
  u_m^{(0)}
  =
  \frac{\displaystyle\int_0^1 (1-r^2)rD^{-1}(r)\,\dd r}
       {\displaystyle\int_0^1 rD^{-1}(r)\,\dd r}
  .
\label{eq:um0}
\end{equation}
When \(D\equiv1\), this expression gives \(u_m^{(0)}=1/2\). In general, \eqref{eq:um0} is equivalent to \(u_m^{(0)}=\int_0^1u(r)\pi(r)\,\dd r\), i.e. the particles average the velocity profile with respect to the equilibrium sampling measure.

The first radial correction is expressed through a cell function \(G(r)\), defined by
\begin{equation}
  D(r)C^{(1)}
  =
  \Per\,G(r)\frac{\partial C_m}{\partial\xi}.
\label{eq:G-definition}
\end{equation}
Substitution into \eqref{eq:C1-equation} gives
\begin{equation}
  \frac1r\frac{\dd}{\dd r}
  \left(r\frac{\dd G}{\dd r}\right)
  =
  \frac{u(r)-u_m^{(0)}}{D(r)}
  ,
\label{eq:cell}
\end{equation}
with boundary conditions
\begin{equation}
  rG'(r)\to0\quad(r\to0),\qquad G'(1)=0.
\label{eq:G-bc}
\end{equation}
The additive constant in \(G\) does not affect \(G'\) or \(\kappa\). A unique normalization may be imposed by
\begin{equation}
  \int_0^1 rG(r)D^{-1}(r)\,\dd r=0 .
\end{equation}
Integrating \eqref{eq:cell} gives
\begin{equation}
  G'(r)=
  \frac1r\int_0^r
  s\,\frac{u(s)-u_m^{(0)}}{D(s)}\,\dd s .
\label{eq:Gprime}
\end{equation}
If the cumulative velocity imbalance is defined as
\begin{equation}
  F(r)=\int_0^r
  s\,\frac{u(s)-u_m^{(0)}}{D(s)}\,\dd s ,
\end{equation}
then \(G'(r)=F(r)/r\). Thus \(G'(r)\) is the radial gradient of the first corrector generated by the cumulative weighted velocity imbalance from the centreline to radius \(r\). The wall condition \(G'(1)=0\) is equivalent to the solvability condition \eqref{eq:um0}.

\subsection{Cross-diffusion correction and Taylor coefficient}

The \(O(\epsilon^2)\) solvability condition separates the roles of the tensor components. The radial component \(D_{rr}\) determines the invariant measure and the first cell problem, and hence controls the leading \(\Pe^2\)-scaled Taylor coefficient. The off-diagonal component \(A=D_{rz}\) contributes to the drift term proportional to \(\partial_\xi C_m\), producing an \(O(\Pe^{-1})\) correction to the laboratory-frame migration speed. The axial component \(B=D_{zz}\) gives an \(O(1)\) direct axial diffusivity and is subleading relative to \(\kappa\Pe^2\).

The conservative form of the axial flux determines the cross-diffusion contribution. Using \eqref{eq:closed-fluxes}, \(J_z=\Pe uc-(Ac)_r-Bc_z\), the \(O(\epsilon^2)\) solvability problem contains the Taylor-dispersion term generated by \(C^{(1)}\), the drift correction generated by \(A=D_{rz}\), and the boundary contribution from the second radial correction \(C^{(2)}\). Before radial averaging, the relevant terms can be arranged as
\begin{equation}
\begin{split}
  \frac{1}{D}\frac{\partial C_m}{\partial T}
  &=
  -\Per^2\frac{[u(r)-u_m^{(0)}]G(r)}{D(r)}
  \frac{\partial^2 C_m}{\partial\xi^2}
  +\Per u_m^{(1)}\frac{1}{D(r)}
  \frac{\partial C_m}{\partial\xi}\\
  &\quad
  +\left[
  \left(\frac{A}{D}\right)'
  +\frac1r\left(r\frac{A}{D}\right)'
  \right]\frac{\partial C_m}{\partial\xi}
  +\mathcal L_0 C^{(2)} .
\end{split}
\label{eq:third-order}
\end{equation}
The first term in this expression is proportional to \(\partial_{\xi\xi}C_m\) and determines axial spreading. The second and third terms are proportional to \(\partial_\xi C_m\) and determine the next-order correction to the mean speed. The \(O(\epsilon^2)\) wall no-flux condition is
\begin{equation}
  \frac{\dd}{\dd r}\left[D(r)C^{(2)}\right]_{r=1}
  +\frac{A(1)}{D(1)}
  \frac{\partial C_m}{\partial\xi}=0.
\label{eq:C2-bc}
\end{equation}
With \(Q=A/D\), the cross-diffusion terms in \eqref{eq:third-order} contribute
\begin{equation}
  \int_0^1 r\left[
  Q'(r)+\frac1r(rQ(r))'
  \right]\dd r
  =
  2Q(1)-\int_0^1 Q(r)\,\dd r .
\end{equation}
The wall condition \eqref{eq:C2-bc} contributes \(-Q(1)\), leaving \(Q(1)-\int_0^1 Q(r)\,\dd r\). Multiplying \eqref{eq:third-order} by \(r\), integrating over \(0<r<1\), and collecting the coefficient of \(\partial_\xi C_m\) therefore gives
\begin{equation}
  u_A
  =
  -
  \frac{
  \dfrac{A(1)}{D(1)}
  -\displaystyle\int_0^1\frac{A(r)}{D(r)}\,\dd r
  }{
  \displaystyle\int_0^1 rD^{-1}(r)\,\dd r
  } .
\label{eq:uA}
\end{equation}
Here
\begin{equation}
  u_A=\Per u_m^{(1)}
\end{equation}
is the \(\Pe\)-scaled correction to the laboratory-frame migration speed. The mean speed in the laboratory coordinate is therefore
\begin{equation}
  u_m=u_m^{(0)}+\Pe^{-1}u_A+O(\Pe^{-2}).
\label{eq:um-expansion-lab}
\end{equation}

Collecting the remaining terms proportional to \(\partial_{\xi\xi}C_m\) yields the slow-time diffusion equation
\begin{equation}
  \frac{\partial C_m}{\partial T}
  =
  \kappa\Per^2
  \frac{\partial^2 C_m}{\partial\xi^2},
\label{eq:Cm-slow}
\end{equation}
where
\begin{equation}
  \kappa
  =
  -
  \frac{\displaystyle\int_0^1
  r\,\frac{[u(r)-u_m^{(0)}]G(r)}{D(r)}\,\dd r}
  {\displaystyle\int_0^1 rD^{-1}(r)\,\dd r}.
\label{eq:kappa-G}
\end{equation}
Integration by parts using \eqref{eq:cell} gives a positive form for \(\kappa\). Specifically, multiplying \eqref{eq:cell} by \(rG\) and integrating yields
\begin{equation}
  \int_0^1 G\frac{\dd}{\dd r}(rG')\,\dd r
  =
  \int_0^1 rG\frac{u-u_m^{(0)}}{D}\,\dd r.
\end{equation}
After integration by parts on the left, the boundary term \([rGG']_0^1\) vanishes by \eqref{eq:G-bc}, so
\begin{equation}
  -\int_0^1 r[G'(r)]^2\,\dd r
  =
  \int_0^1 rG(r)\frac{u(r)-u_m^{(0)}}{D(r)}\,\dd r.
\end{equation}
Substitution into \eqref{eq:kappa-G} gives
\begin{equation}
  \kappa
  =
  \frac{\displaystyle\int_0^1 r[G'(r)]^2\,\dd r}
       {\displaystyle\int_0^1 rD^{-1}(r)\,\dd r}
  .
\label{eq:kappa}
\end{equation}
This form makes the positivity of \(\kappa\) explicit. The dispersion-energy density
\begin{equation}
  I_\kappa(r)=
  \frac{r[G'(r)]^2}
       {\displaystyle\int_0^1 sD^{-1}(s)\,\dd s}
\label{eq:Ikappa}
\end{equation}
gives the contribution of each radius to the leading Taylor coefficient. Substitution of \eqref{eq:Gprime} further gives the double-integral representation
\begin{equation}
  \kappa
  =
  \frac{
  \displaystyle
  \int_0^1 \frac1r
  \left[
  \int_0^r s\,\frac{(1-s^2)-u_m^{(0)}}{D(s)}\,\dd s
  \right]^2\dd r
  }
  {
  \displaystyle
  \int_0^1 rD^{-1}(r)\,\dd r
  } .
\label{eq:kappa-double}
\end{equation}
Returning to the laboratory coordinate, the long-time one-dimensional equation is
\begin{equation}
  \frac{\partial C_m}{\partial t}
  +\Pe u_m\frac{\partial C_m}{\partial z}
  =
  \left[\kappa\Pe^2+K_{\rm dir}+o(1)\right]
  \frac{\partial^2 C_m}{\partial z^2}
  .
\label{eq:effective1d}
\end{equation}
The direct axial contribution associated with the leading radial equilibrium is
\begin{equation}
  K_{\rm dir}
  =
  \frac{\displaystyle\int_0^1 rB(r)D^{-1}(r)\,\dd r}
       {\displaystyle\int_0^1 rD^{-1}(r)\,\dd r},
  \qquad B(r)=D_{zz}(r).
\label{eq:Kdir}
\end{equation}
When the dispersion coefficient is defined through the variance growth rate scaled by \(2\Pe^2\), \(K_{\rm dir}/\Pe^2\) is a higher-order correction, while \(K_{\rm dir}\) remains part of the unscaled axial variance. In \eqref{eq:effective1d}, \(o(1)\) denotes terms that vanish as \(\Pe\to\infty\) at fixed \(\Per\) and \(p\), equivalently under the ordering \(\Per/\Pe\to0\). The spherical limit gives
\begin{equation}
  D\equiv1,\qquad A=0,\qquad
  u_m=\frac12,\qquad
  u_A=0,\qquad
  \kappa_s=\frac{1}{192}.
\label{eq:spherical-limit}
\end{equation}
This recovers the classical tube Poiseuille coefficient \citep{taylor1953,aris1956}.

\section{Long-time transport coefficients and radial mechanisms}
\label{sec:long-time-coefficients}

\subsection{Mean speed and Taylor dispersion}

\begin{figure}
  \centering
  \includegraphics[width=0.98\linewidth]{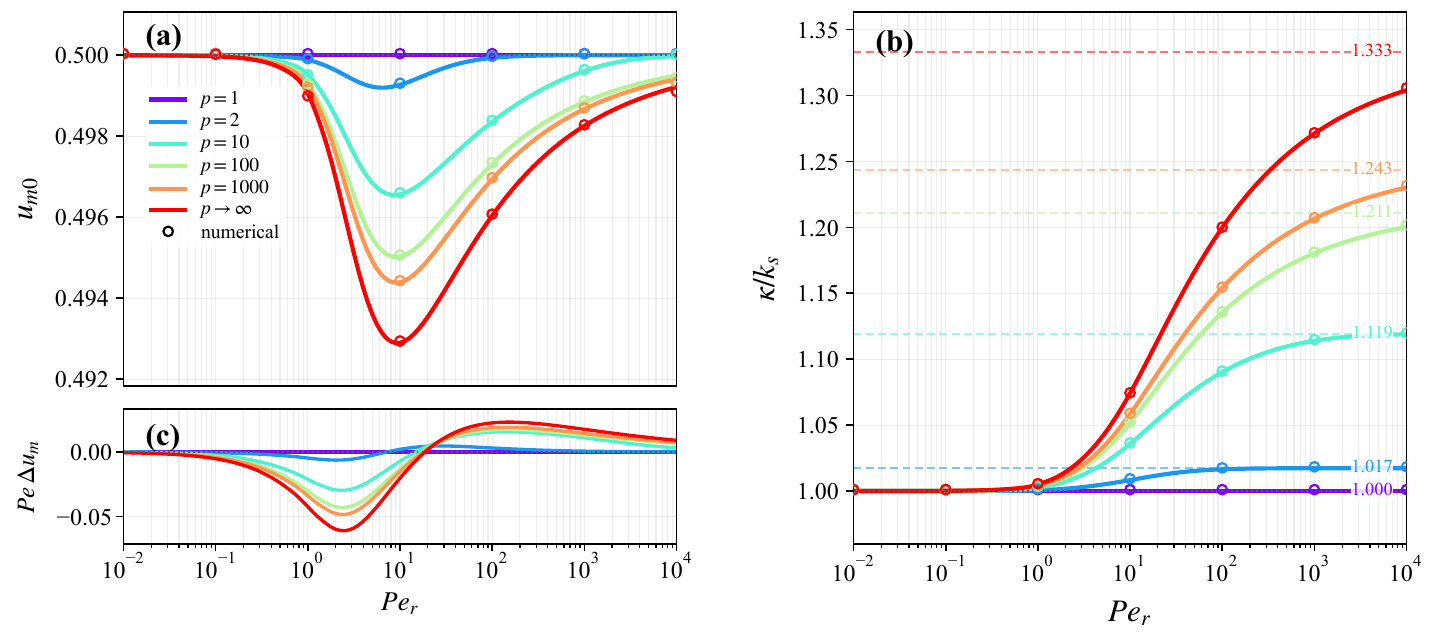}
  \caption{Variation of long-time mean speed and Taylor dispersion coefficient with \(\Per\). Panel (a) shows the leading-order mean speed \(u_m^{(0)}\). Panel (b) shows the Taylor dispersion coefficient normalized by the classical spherical value \(\kappa_s=1/192\); the dashed lines indicate estimated limits as \(\Per\to\infty\). Panel (c) shows the \(O(\Pe^{-1})\) mean-speed correction coefficient \(u_A=\Pe\Delta u_m\) induced by cross-diffusion \(D_{rz}\). Open symbols are long-time coefficients fitted from direct evolution of the full axisymmetric tensorial equation.}
  \label{fig:mean-speed-kappa}
\end{figure}

Section~\ref{sec:taylor-aris-reduction} reduces the full tensorial problem to three radial quantities: the invariant measure in \eqref{eq:pi-measure}, the \(\Pe^2\)-scaled Taylor coefficient through the energy form \eqref{eq:kappa}, and the cross-diffusion speed correction in \eqref{eq:uA}. Their dependence on \(\Per\) and aspect ratio \(p\) reveals how local orientational alignment changes macroscopic migration and dispersion, giving the circular-tube analogue of the passive-rod mechanism identified in planar Poiseuille flow \citep{kumar2021,khair2022} (Figure~\ref{fig:mean-speed-kappa}).

The spherical limit gives the horizontal baselines \(u_m^{(0)}=1/2\), \(\kappa/\kappa_s=1\) and \(u_A=0\). For \(p>1\), \(u_m^{(0)}\) varies non-monotonically with \(\Per\). At intermediate \(\Per\), the reduction of \(D_{rr}\) is strongest in the outer part of the tube, where the Poiseuille velocity is smaller. The invariant measure \(rD^{-1}(r)\,\dd r\) therefore gives more weight to slower fluid layers, and \(u_m^{(0)}\) falls below \(1/2\). For \(p=1000\), the minimum is approximately \(0.4944\); as \(p\to\infty\), the minimum is approximately \(0.4929\). At larger \(\Per\), the aligned region occupies most of the cross-section, so \(D_{rr}(r)\) becomes nearly uniform except near the low-shear centreline. The invariant measure then approaches the area measure again, and \(u_m^{(0)}\) returns towards \(1/2\).

The enhancement of Taylor dispersion follows from the same cell problem. At small \(\Per\), the local shear is weak across the cross-section, the orientation distribution remains close to isotropic, and \(\kappa/\kappa_s\) is close to unity. As \(\Per\) increases, streamwise alignment in the outer region decreases \(D_{rr}\), slows transverse exchange, and allows velocity deviations to persist longer during radial mixing, strengthening the response of the cell problem. At \(\Per=10^4\), the enhancement is approximately \(1.7\%\) for \(p=2\), \(23.0\%\) for \(p=1000\), and \(30.4\%\) in the infinitely slender rod limit.

Following the maximum-enhancement normalization used for planar rods \citep{kumar2021}, the hypothetically fully aligned state provides a useful theoretical reference scale. If all rods were aligned with the flow, \(D_{rr}=D_\perp/\Dbar\) would be spatially uniform and the tube Taylor coefficient would satisfy \(\kappa=\kappa_s/(D_\perp/\Dbar)\). Thus
\begin{equation}
  \frac{\kappa_m}{\kappa_s}=\frac{\Dbar}{D_\perp}.
\label{eq:kappa-upper}
\end{equation}
Here \(\kappa_m\) denotes the fully aligned reference coefficient.
For the slender limit \(D_\parallel\to2D_\perp\), this expression gives \(\kappa_m/\kappa_s\to4/3\). At the largest value shown, \(\Per=10^4\), the \(p\to\infty\) curve reaches \(\kappa/\kappa_s\simeq1.304\). The remaining gap from \(4/3\) reflects finite-\(\Per\) orientational diffusion and the low-shear centreline region, where the orientation distribution stays broader. This reference scale motivates the normalized enhancement factor
\begin{equation}
  E(\Per,p)=
  \frac{\kappa/\kappa_s-1}
       {\kappa_m/\kappa_s-1},
\label{eq:enhancement-factor}
\end{equation}
which measures the fraction of the fully aligned enhancement attained at the current shear strength.

\begin{figure}
  \centering
  \includegraphics[width=0.78\linewidth]{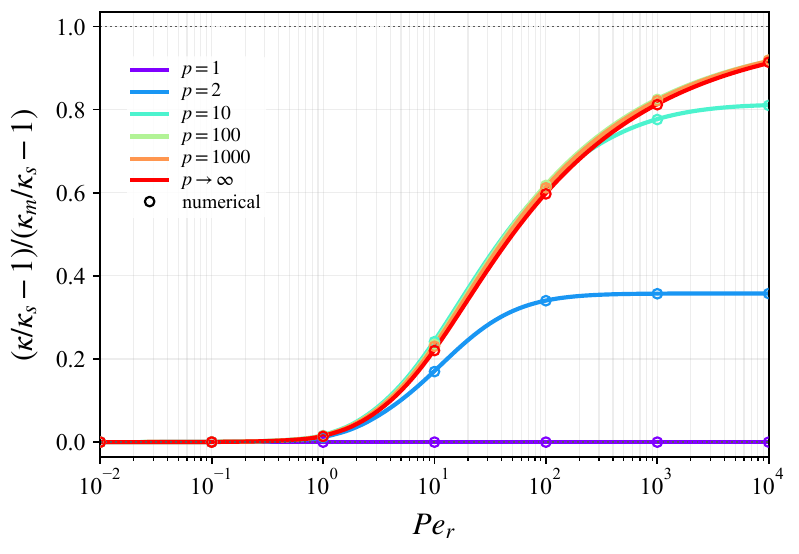}
  \caption{Variation of the normalized Taylor-dispersion enhancement factor \(E\) with \(\Per\). Solid curves are calculated from the long-time theory in Figure~\ref{fig:mean-speed-kappa}. Open symbols are fitted results from the full axisymmetric tensorial equation at \(\Pe=10^5\), with the numerical baseline bias removed using a spherical case at the same \(\Pe\). The case \(p=1\) has no available enhancement interval and is shown as the spherical baseline \(E=0\).}
  \label{fig:enhancement-collapse}
\end{figure}

With the normalization in \eqref{eq:enhancement-factor}, the curves for \(p\gtrsim10\) nearly collapse over a broad range of \(\Per\). The enhancement process for slender rods is governed primarily by the rotational P{\'e}clet number, with the aspect ratio setting the maximum attainable enhancement. The curve for \(p=2\) lies appreciably below the main curve, indicating stronger residual shape dependence for short rods. For \(p=1000,\Per=10^4\), \(E\simeq0.92\), so the radial response is already close to the fully aligned scale.

The cross-diffusion correction in \eqref{eq:uA} is set by the wall value and the cross-sectional integral of \(D_{rz}\). Their competition makes \(u_A\) non-monotone in \(\Per\) and allows a change of sign. For \(p=1000\), this correction coefficient ranges from approximately \(-4.84\times10^{-2}\) to \(1.92\times10^{-2}\). Its contribution to the actual mean speed is multiplied by \(\Pe^{-1}\), making it smaller than the leading speed in the high-\(\Pe\) limit.

\begin{table}
  \centering
  \caption{Summary of the parameter scan in Figure~\ref{fig:mean-speed-kappa}. The column \(u_{m,\min}^{(0)}\) gives the minimum over \(\Per\in[10^{-2},10^4]\); \(u_m^{(0)}\) and \(\kappa/\kappa_s\) are reported at \(\Per=10^4\); the \(u_A\) range is taken over the same \(\Per\) interval.}
  \label{tab:fig5-summary}
  \begin{tabular}{ccccc}
    \toprule
    \(p\) & \(u_{m,\min}^{(0)}\) & \(u_m^{(0)}\) & \(\kappa/\kappa_s\) & \(u_A\) range \\
    \midrule
    \(1\) & 0.5000 & 0.5000 & 1.000 & \(0\) \\
    \(2\) & 0.4992 & 0.5000 & 1.017 & \([-0.0063,\ 0.0047]\) \\
    \(10\) & 0.4965 & 0.5000 & 1.119 & \([-0.0297,\ 0.0156]\) \\
    \(100\) & 0.4950 & 0.4995 & 1.200 & \([-0.0430,\ 0.0174]\) \\
    \(1000\) & 0.4944 & 0.4994 & 1.230 & \([-0.0484,\ 0.0192]\) \\
    \(\infty\) & 0.4929 & 0.4992 & 1.304 & \([-0.0610,\ 0.0231]\) \\
    \bottomrule
  \end{tabular}
\end{table}

\subsection[Radial tensor profiles induced by q=Per r]{Radial tensor profiles induced by \(q=\Per r\)}

\begin{figure}
  \centering
  \includegraphics[width=0.98\linewidth]{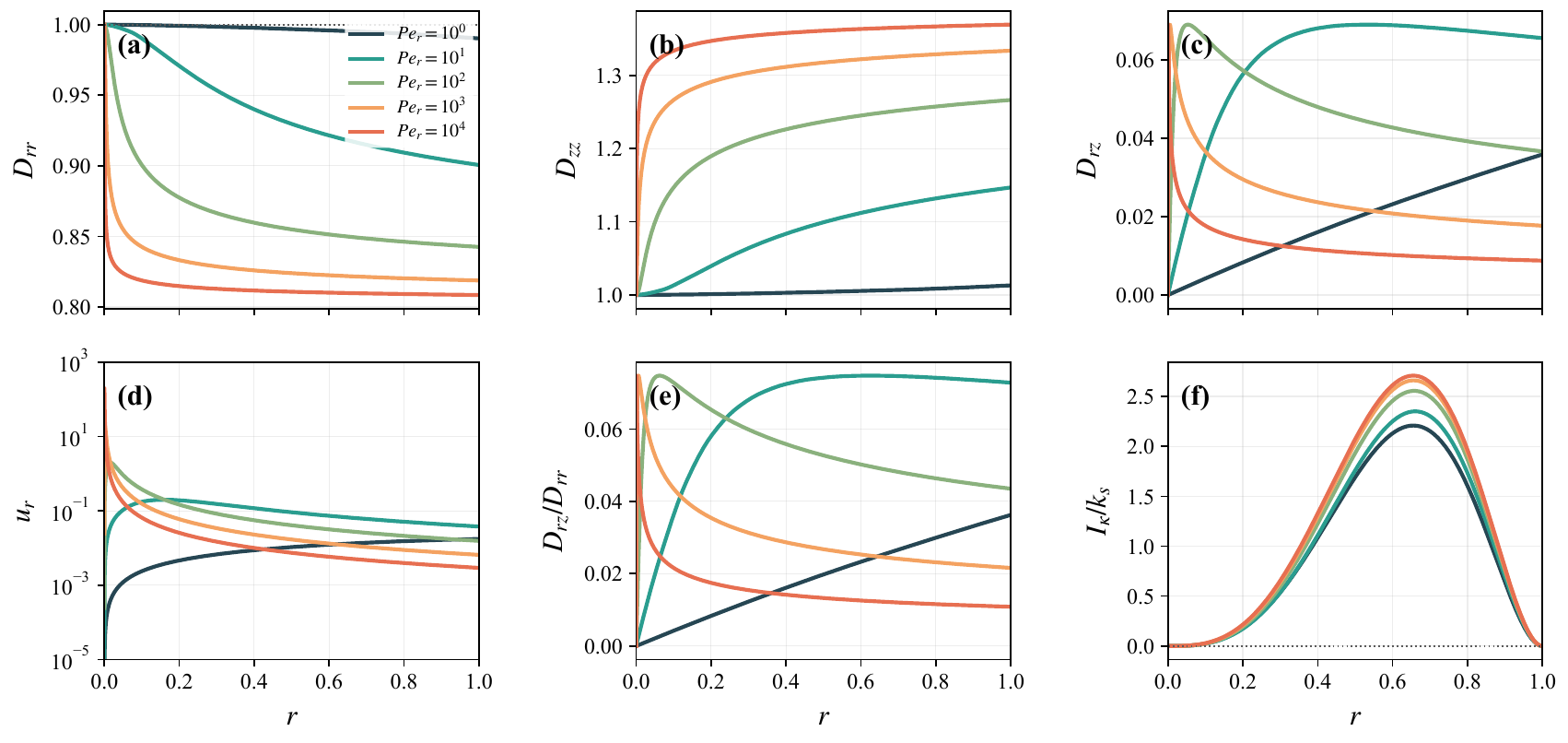}
  \caption{Radial tensor profiles and dispersion contribution density for fixed \(p=1000\). Panel (a) shows \(D_{rr}(r)\). Panel (b) shows \(D_{zz}(r)\). Panel (c) shows \(D_{rz}(r)\). Panel (d) shows \(v_D^+(r)=\max[-D_{rr}'(r),0]\), the positive part of the statistical drift associated with radially inhomogeneous \(D_{rr}\). Panel (e) shows \(D_{rz}/D_{rr}\). Panel (f) shows the Taylor-dispersion contribution density \(I_\kappa/\kappa_s\). Colours correspond to \(\Per=10^0,10^1,10^2,10^3,10^4\).}
  \label{fig:radial-tensor-profiles}
\end{figure}

Transverse mixing weakens as \(\Per\) increases because orientational alignment strengthens in the outer shear region. The radial diffusivity \(D_{rr}(r)\) decreases from the spherical reference value \(1\) over a broad radial interval (Figure~\ref{fig:radial-tensor-profiles}(a)). Because the leading radial equilibrium gives \(C^{(0)}\propto D_{rr}^{-1}\), and the cell-problem forcing contains \([u(r)-u_m^{(0)}]/D_{rr}(r)\), low-\(D_{rr}\) regions affect both the mean speed and the Taylor response.

The contribution density \(I_\kappa(r)\), defined by \eqref{eq:Ikappa}, concentrates the dominant share of \(\kappa\) in the middle-to-outer part of the tube (Figure~\ref{fig:radial-tensor-profiles}(f)), where both the velocity contrast and the cell-function slope are substantial. The excess relative to the spherical case is isolated in Figure~\ref{fig:kappa-mechanism}.

The off-diagonal coefficient \(D_{rz}\) is appreciable over the intermediate-shear region (Figure~\ref{fig:radial-tensor-profiles}(c,e)). In the full equation it enters both
\begin{equation}
  J_r=\cdots-D_{rz}c_z,\qquad
  J_z=\cdots-\partial_r(D_{rz}c).
\end{equation}
In the long-wave reduction, its leading contribution in the asymptotic ordering is the \(O(\Pe^{-1})\) mean migration-speed correction \eqref{eq:uA}. Its influence on the leading Taylor dispersion coefficient scaled by \(\Pe^2\) enters at higher order. The agreement between coefficients fitted from the full tensorial equation and the leading theory in Figure~\ref{fig:mean-speed-kappa}(b) supports the asymptotic ordering: the \(\Pe^2\)-scaled dispersion is controlled primarily by \(D_{rr}(r)\) through the invariant measure and the cell problem. The cross-diffusion coefficient \(D_{rz}\) appears at leading order in the speed correction \(u_A\) (Figure~\ref{fig:mean-speed-kappa}(c)). The coefficient \(D_{zz}\) corresponds to the direct axial diffusion term in the full equation (Figure~\ref{fig:radial-tensor-profiles}(b)). It contributes \(O(1)\) to the unscaled axial variance and is a higher-order correction in the leading Taylor coefficient scaled by \(\Pe^2\).

These profiles separate the asymptotic roles of the tensor components. The leading \(\Pe^2\)-scaled Taylor coefficient is controlled by \(D_{rr}\) through the invariant measure and the cell problem. The off-diagonal coefficient \(D_{rz}\) generates the \(O(\Pe^{-1})\) speed correction, and \(D_{zz}\) contributes to the \(O(1)\) direct axial diffusivity. A scalar diffusivity replacement would collapse these distinct operators into a single coefficient and lose their separate contributions.

\subsection{Radial origin of dispersion enhancement}

\begin{figure}
  \centering
  \includegraphics[width=\linewidth]{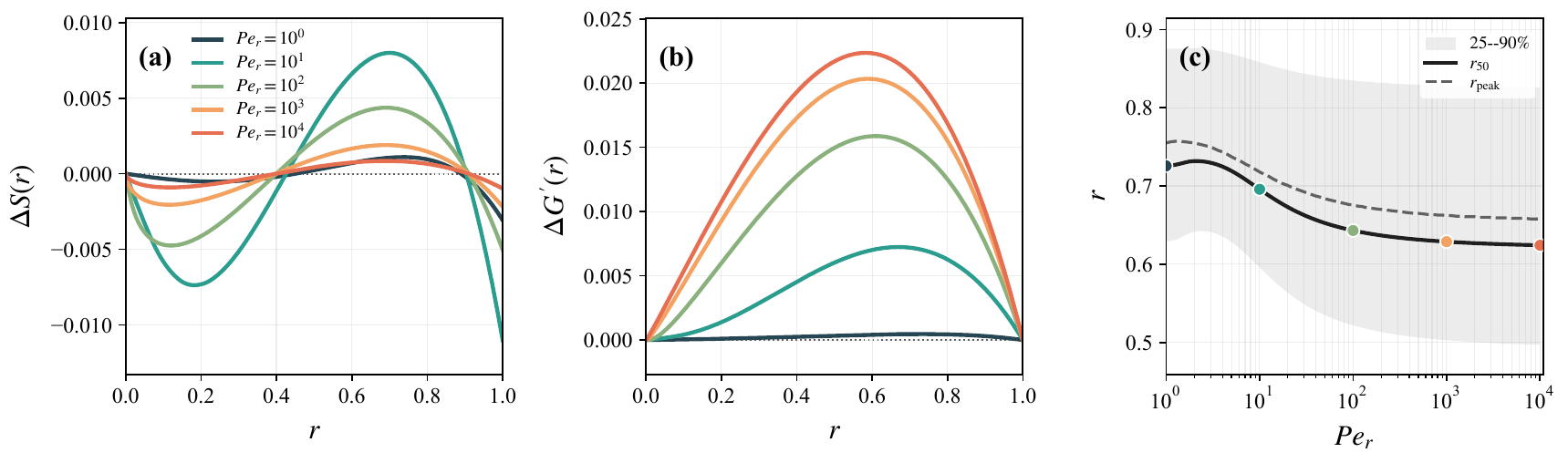}
  \caption{Radial mechanism of the dispersion change at fixed \(p=1000\). Panel (a) shows the change in source relative to the spherical reference, \(\Delta S(r)=S(r)-S_s(r)\). Panel (b) shows the change in cell-function slope, \(\Delta G'(r)=G'(r)-G_s'(r)\). Panel (c) shows the radial origin of the excess dispersion contribution. The grey band indicates the \(25\%\)--\(90\%\) interval of cumulative excess contribution, the black solid line is the median radius \(r_{50}\), and the grey dashed line is the radius \(r_{\rm peak}\) at which the local excess contribution density is maximal.}
  \label{fig:kappa-mechanism}
\end{figure}

To identify the spatial origin of the enhancement in \(\kappa\), fix \(p=1000\) and compare the cell problems for rod-like and spherical particles. The source term, corrector slope and energy density provide a radial decomposition of the excess Taylor response. The normalized source is
\begin{equation}
  S(r)=
  \frac{rD^{-1}(r)[u(r)-u_m^{(0)}]}
       {\displaystyle\int_0^1 sD^{-1}(s)\,\dd s},
\label{eq:S-source}
\end{equation}
and the spherical reference source
\begin{equation}
  S_s(r)=2r\left(\frac12-r^2\right).
\end{equation}
By \eqref{eq:um0}, \(\int_0^1S(r)\,\dd r=0\). The source \(S(r)\) is the weighted velocity-deviation density associated with the invariant measure \(rD^{-1}(r)\,\dd r\); its positive and negative parts identify radial layers moving faster or slower than the particle mean speed. Also let
\begin{equation}
  G_s'(r)=\frac{r(1-r^2)}4
\end{equation}
be the cell-function slope for the spherical tube problem. The slope \(G'(r)\) is the radial integral of the weighted source, up to the factor \(I_0/r\). The quantity \(r[G'(r)]^2/I_0\) in \eqref{eq:Ikappa} is the positive energy density entering the Taylor coefficient. The cumulative excess dispersion is
\begin{equation}
  \Delta K(r)
  =
  \frac{1}{\kappa_s}
  \int_0^r
  \left[
  \frac{s[G'(s)]^2}{\displaystyle\int_0^1 \rho D^{-1}(\rho)\,\dd\rho}
  -2s\left(\frac{s(1-s^2)}4\right)^2
  \right]\dd s .
\label{eq:DeltaK}
\end{equation}
Then \(\Delta K(1)=(\kappa-\kappa_s)/\kappa_s\). The radii \(r_\alpha\) in Figure~\ref{fig:kappa-mechanism}(c) are defined by
\begin{equation}
  \Delta K(r_\alpha)=\alpha\,\Delta K(1),
  \qquad \alpha=0.25,\ 0.50,\ 0.90 .
\end{equation}
For the cases shown, the cumulative excess is monotone over the interval used for this diagnostic.

Shear-induced variation in \(D(r)\) reshapes the source term (Figure~\ref{fig:kappa-mechanism}(a)). At larger \(\Per\), \(\Delta S\) is negative near the centre, positive at middle to outer radii, and then decreases near the wall. Because \(G'(r)\) is the radial integral of this source, a positive shift of \(S(r)\) in the middle-to-outer annulus increases the magnitude of the corrector slope over the same region. Since the Taylor coefficient is proportional to \(\int_0^1 r[G'(r)]^2\,\dd r\), this redistribution increases \(\kappa\). For \(p=1000\), \(\max G'\) increases from \(0.0966\) at \(\Per=10^0\) to \(0.1186\) at \(\Per=10^4\), while \(\kappa/\kappa_s\) increases from \(1.004\) to \(1.230\).

The radial origin of the additional dispersion moves inward as alignment strengthens (Figure~\ref{fig:kappa-mechanism}(c)). As \(\Per\) increases from \(10^0\) to \(10^4\), the peak radius of the excess contribution density moves from approximately \(0.755\) to \(0.658\), and the median contribution radius moves from approximately \(0.726\) to \(0.624\). At \(\Per=10^4\), the \(25\%\)--\(90\%\) cumulative contribution interval is approximately \(0.497<r<0.826\). The leading additional Taylor dispersion is therefore generated by a broad annulus spanning middle to outer radii.

\begin{table}
  \centering
  \caption{Summary of the mechanism quantities for \(p=1000\) in Figure~\ref{fig:kappa-mechanism}.}
  \label{tab:fig7-summary}
  \begin{tabular}{ccccccc}
    \toprule
    \(\Per\) & \(D_{rr}(1)\) & \(\kappa/\kappa_s\) & \(\Delta\kappa/\kappa_s\) & \(r_{\rm peak}\) & \(r_{50}\) & \(r_{25}\)--\(r_{90}\) \\
    \midrule
    \(10^0\) & 0.990 & 1.004 & 0.004 & 0.755 & 0.726 & 0.630--0.875 \\
    \(10^1\) & 0.901 & 1.058 & 0.058 & 0.718 & 0.696 & 0.596--0.858 \\
    \(10^2\) & 0.843 & 1.154 & 0.154 & 0.676 & 0.643 & 0.522--0.835 \\
    \(10^3\) & 0.819 & 1.207 & 0.207 & 0.662 & 0.629 & 0.503--0.828 \\
    \(10^4\) & 0.808 & 1.230 & 0.230 & 0.658 & 0.624 & 0.497--0.826 \\
    \bottomrule
  \end{tabular}
\end{table}

The radial decomposition isolates the annular origin of the enhancement. The reduction of \(D_{rr}\) shifts the weighted velocity-deviation source toward the middle-to-outer part of the tube. This redistribution increases the corrector slope \(G'(r)\) over the same region, and the positive energy
\begin{equation}
  \int_0^1 r[G'(r)]^2\,\dd r
\end{equation}
therefore exceeds its spherical counterpart. The Taylor-dispersion enhancement is generated primarily by a broad annulus centred away from both the centreline and the wall.

\section{Full-time radial spectral representation}
\label{sec:spectral-representation}

Finite-time packets approach the Taylor--Aris regime through the decay of radial non-equilibrium modes. A spectral representation makes this relaxation structure explicit by projecting the initial radial distribution onto the modes of the radial mixing operator. The modal amplitudes record how the packet samples the velocity profile during relaxation and how the accumulated response recovers the Taylor--Aris coefficient. This formulation is useful for comparing injections and background velocity profiles, because the radial physics is carried by a fixed Sturm--Liouville operator and the shear enters through a velocity matrix. Direct integration of the full tensorial equation provides the transient concentration field and the high-fidelity reference used for validation below.

\subsection{Reduced-order propagation equation}

The Taylor--Aris coefficient derived above is the long-time response of the fundamental radial mode, after all transient radial modes have relaxed. Finite-time packets in experiments and simulations can retain memory of their injection profile, so the approach to the Taylor limit must track these decaying radial modes \citep{vedel2012,vedel2014,guan2024}. We use the reduced propagation model controlled by the same radial operator that appears in the long-time cell problem:
\begin{equation}
  c_t+\Pe u(r)c_z
  =
  \frac1r\frac{\partial}{\partial r}
  \left[
  r\frac{\partial}{\partial r}\bigl(D(r)c\bigr)
  \right],
\label{eq:alltime-model}
\end{equation}
where \(D(r)=D_{rr}(\Per r)\). This \(D_{rr}\)-dominated full-time model isolates radial mixing, velocity sampling and their coupling to axial shear. The cross-diffusive fluxes involving \(D_{rz}\) and the direct axial diffusion from \(D_{zz}\) are retained in the full tensorial equation used for validation. In the present spectral reduction they enter beyond the leading \(\Pe^2\)-scaled dispersion coefficient under the high-\(\Pe\) ordering of Section~\ref{sec:taylor-aris-reduction}. The model therefore interprets the modal route from radial non-equilibrium to the Taylor limit, with its range checked against the full equation. The wall and axis conditions are
\begin{equation}
  \partial_r(Dc)\big|_{r=1}=0,\qquad
  r\,\partial_r(Dc)\to0\quad(r\to0).
\end{equation}

The weighted concentration
\begin{equation}
  w(r,z,t)=D(r)c(r,z,t),
\label{eq:w-def}
\end{equation}
places \eqref{eq:alltime-model} in the form
\begin{equation}
  w_t+\Pe u(r)w_z=\mathcal L_r w,
  \qquad
  \mathcal L_r w=
  D(r)\frac1r\frac{\partial}{\partial r}
  \left(r\frac{\partial w}{\partial r}\right).
\label{eq:w-equation}
\end{equation}
With the weighted inner product
\begin{equation}
  (f,g)_D=\int_0^1\frac{r}{D(r)}f(r)g(r)\,\dd r
\label{eq:weighted-inner-product}
\end{equation}
\(\mathcal L_r\) is a self-adjoint negative operator. Integration by parts using the boundary conditions gives
\begin{equation}
  (\mathcal L_r f,g)_D
  =
  -\int_0^1 r f'(r)g'(r)\,\dd r
  =
  (f,\mathcal L_r g)_D .
\label{eq:self-adjoint}
\end{equation}

\subsection{\texorpdfstring{Sturm--Liouville modes and Fourier propagator}{Sturm-Liouville modes and Fourier propagator}}

The self-adjoint operator supplies an orthonormal set of radial relaxation modes. They are determined by
\begin{equation}
  -\mathcal L_r\phi_n=\lambda_n\phi_n,
\label{eq:eig-op}
\end{equation}
or equivalently
\begin{equation}
  -\frac{\dd}{\dd r}\left(r\frac{\dd\phi_n}{\dd r}\right)
  =
  \lambda_n\frac{r}{D(r)}\phi_n .
\label{eq:spectral-eig}
\end{equation}
Regularity, no flux and normalization impose
\begin{equation}
  r\phi_n'(r)\to0\quad(r\to0),\qquad
  \phi_n'(1)=0,\qquad
  (\phi_m,\phi_n)_D=\delta_{mn}.
\label{eq:eig-bc}
\end{equation}
Here \(\lambda_0=0\), \(\phi_0\) is the constant mode, and the modes \(n\ge1\) describe the decay of radial non-equilibrium.

Expanding the weighted concentration in these modes,
\begin{equation}
  w(r,z,t)=\sum_{n=0}^{\infty}a_n(z,t)\phi_n(r)
\label{eq:w-expansion}
\end{equation}
and projecting \eqref{eq:w-equation} onto \(\phi_n\) using \((\cdot,\cdot)_D\), gives
\begin{equation}
  \frac{\partial a_n}{\partial t}
  +\Pe\sum_{m=0}^{\infty}U_{nm}
  \frac{\partial a_m}{\partial z}
  =
  -\lambda_n a_n,
\label{eq:modal-pde}
\end{equation}
where the velocity matrix is
\begin{equation}
  U_{nm}=
  \int_0^1
  \frac{r}{D(r)}u(r)\phi_n(r)\phi_m(r)\,\dd r .
\label{eq:velocity-matrix}
\end{equation}
Axial Fourier transformation with axial wavenumber \(k\) converts the modal equations into
\begin{equation}
  \frac{\dd\widehat{\bm a}}{\dd t}
  =
  -\left(\Lambda+i k\Pe U\right)\widehat{\bm a},
  \qquad
  \Lambda=\operatorname{diag}(\lambda_0,\lambda_1,\ldots).
\end{equation}
Therefore
\begin{equation}
  \widehat{\bm a}(k,t)
  =
  \exp\!\left[-\left(\Lambda+i k\Pe U\right)t\right]
  \widehat{\bm a}(k,0)
  .
\label{eq:fourier-propagator}
\end{equation}
For an arbitrary axisymmetric initial concentration \(c_0(r,z)\), projection gives the initial modal amplitudes
\begin{equation}
  a_n(z,0)=\int_0^1 r\,\phi_n(r)c_0(r,z)\,\dd r
\label{eq:initial-an}
\end{equation}
and the concentration field is reconstructed as
\begin{equation}
  c(r,z,t)=\frac{1}{D(r)}
  \sum_{n=0}^{\infty}a_n(z,t)\phi_n(r)
\label{eq:c-reconstruction}
\end{equation}
after evolution.

\subsection{Moment equations and long-time limit}

The axial variance can be propagated directly through modal moments, following the moment-based treatment of Taylor--Aris transients \citep{aris1956,vedel2012,vedel2014}. Suppose the initial condition has the form
\begin{equation}
  c(r,z,0)=h(r)F_0(z),
  \qquad
  \int zF_0(z)\,\dd z=0,
  \qquad
  \frac{\int z^2F_0(z)\,\dd z}{\int F_0(z)\,\dd z}
  =\sigma_{z0}^2,
\end{equation}
with initial radial projection
\begin{equation}
  b_n=\int_0^1 r h(r)\phi_n(r)\,\dd r .
\label{eq:initial-projection}
\end{equation}
The first three axial modal moments are
\begin{equation}
  m_n^{(j)}(t)=\int z^j a_n(z,t)\,\dd z,\qquad j=0,1,2.
\end{equation}
Let \(\bm m^{(j)}\) denote the vector with entries \(m_n^{(j)}\).
Multiplying \eqref{eq:modal-pde} by \(z^j\), integrating over \(z\), and integrating the axial derivative by parts gives the closed moment system
\begin{equation}
  \frac{\dd}{\dd t}
  \begin{bmatrix}
    \bm m^{(0)}\\
    \bm m^{(1)}\\
    \bm m^{(2)}
  \end{bmatrix}
  =
  \begin{bmatrix}
    -\Lambda & 0 & 0\\
    \Pe U & -\Lambda & 0\\
    0 & 2\Pe U & -\Lambda
  \end{bmatrix}
  \begin{bmatrix}
    \bm m^{(0)}\\
    \bm m^{(1)}\\
    \bm m^{(2)}
  \end{bmatrix},
\label{eq:moment-system}
\end{equation}
with initial values
\begin{equation}
  \bm m^{(0)}(0)=\bm b,\qquad
  \bm m^{(1)}(0)=0,\qquad
  \bm m^{(2)}(0)=\sigma_{z0}^2\bm b .
\end{equation}
The mean axial position and variance of the cross-sectional concentration packet are obtained from the zero mode:
\begin{equation}
  \avg{z}(t)=\frac{m_0^{(1)}(t)}{b_0},
  \qquad
  \sigma_z^2(t)=
  \frac{m_0^{(2)}(t)}{b_0}
  -\left[\frac{m_0^{(1)}(t)}{b_0}\right]^2 .
\label{eq:spectral-moments}
\end{equation}

The long-time limit of the spectral model recovers the Taylor--Aris coefficients derived above. The zero-mode velocity-matrix element is
\begin{equation}
  u_m^{(0)}=U_{00},
\end{equation}
identical to \eqref{eq:um0}. Non-zero modes are excited by the velocity deviation and decay at rates \(\lambda_n\), giving the spectral sum
\begin{equation}
  \kappa_\infty
  =
  \sum_{n=1}^{\infty}
  \frac{U_{0n}U_{n0}}{\lambda_n}
  .
\label{eq:kappa-spectrum}
\end{equation}
Equation~\eqref{eq:kappa-spectrum} is the eigenmode representation of \eqref{eq:kappa}; both describe the same radial response. The formula also gives a spectral interpretation of Taylor dispersion: modes with strong velocity-profile coupling \(U_{0n}U_{n0}\) and slow radial decay \(\lambda_n^{-1}\) contribute most to the long-time axial variance.

\subsection{Validation for different radial initial conditions}

Radial memory is most visible when the injection profile selects different parts of the velocity field before transverse relaxation. The validation therefore uses three packets with the same axial width: a cross-section-filling packet, a centre-enriched packet in the fast, weak-shear core, and a near-wall-enriched packet in the slow, strongly sheared annulus:
\begin{equation}
  c_0(r,z)=h(r)
  \exp\!\left[-\frac{(z-z_0)^2}{2\sigma_{z0}^2}\right],
  \qquad \sigma_{z0}=6 .
\label{eq:fig8-initial}
\end{equation}
The unnormalized radial weights are
\begin{equation}
  \tilde h_{\rm uniform}=1,\qquad
  \tilde h_{\rm center}=\exp\!\left[-\frac{r^2}{2s_r^2}\right],
  \qquad
  \tilde h_{\rm wall}=\exp\!\left[-\frac{(1-r)^2}{2s_r^2}\right],
\end{equation}
with \(s_r=0.25\). Normalization is imposed by
\begin{equation}
  h(r)=
  \frac{
  \left(\displaystyle\int_0^1 r\,\dd r\right)\tilde h(r)}
  {\displaystyle\int_0^1 r\tilde h(r)\,\dd r},
\label{eq:h-normalization}
\end{equation}
so that all three initial conditions have the same cross-sectional mass,
\begin{equation}
  \int_0^1 rh(r)\,\dd r=\frac12 .
\end{equation}

\begin{figure}
  \centering
  \includegraphics[width=\linewidth]{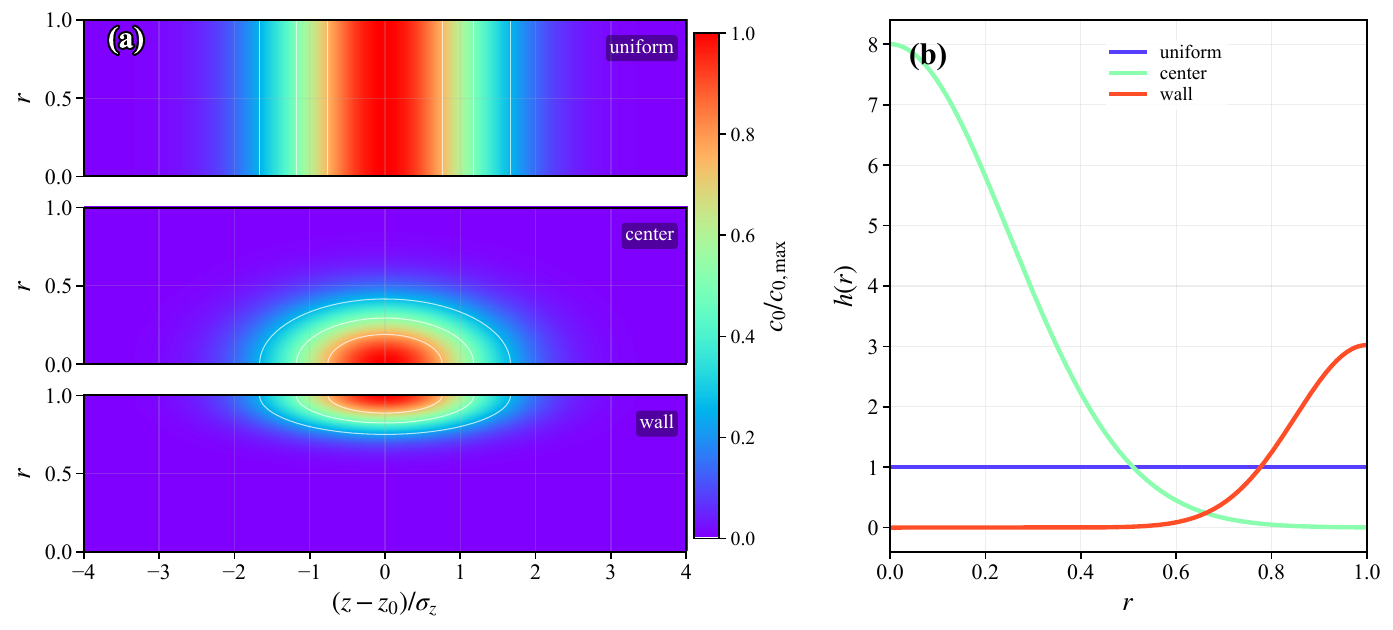}
  \caption{Radial--axial initial conditions used for full-time spectral validation. Panel (a) shows \(c_0(r,z)/c_{0,\max}\), with the horizontal coordinate \((z-z_0)/\sigma_{z0}\); the three rows correspond to uniform, centre-enriched and near-wall-enriched initial conditions. Panel (b) shows the corresponding radial weights \(h(r)\). White contours in panel (a) mark normalized concentrations of \(0.25\), \(0.5\) and \(0.75\). All three \(h(r)\) are normalized by \eqref{eq:h-normalization}.}
  \label{fig:spectral-initial}
\end{figure}

The constructed profiles have equal axial width and cross-sectional mass (Figure~\ref{fig:spectral-initial}). Their radial weights span widely separated transverse sampling states. The uniform initial condition fills the cross-section; the centre-enriched initial condition is concentrated near the centreline, with normalized values \(h(0)\simeq8.003\) and \(h(1)\simeq2.68\times10^{-3}\); the near-wall-enriched initial condition is concentrated close to the wall, with \(h(1)\simeq3.021\) and \(h(0)\simeq6.75\times10^{-10}\). This set starts two cases far from radial equilibrium and directly probes the influence of radial relaxation modes on early variance growth.

The corresponding initial modal projections are given in Table~\ref{tab:initial-modal-projections}. The uniform packet is almost zero-mode dominated, with small non-zero projections caused by the difference between uniform concentration and the invariant radial equilibrium \(c\propto D^{-1}\). The centre- and wall-enriched packets place much larger weight on the first few decaying modes. The first mode also changes sign between centre and wall injection, reflecting that these packets sample opposite sides of the velocity deviation before radial relaxation.

\begin{table}
  \centering
  \caption{Initial modal projections for the profiles in Figure~\ref{fig:spectral-initial}. Entries are signed ratios \(b_n/b_0\), where \(b_n=\int_0^1 r h(r)\phi_n(r)\,\dd r\). Modes are ordered by increasing \(\lambda_n\), and their signs are fixed by \(\int_0^1 r\phi_n(r)\,\dd r>0\).}
  \label{tab:initial-modal-projections}
  \begin{tabular}{ccccccc}
    \toprule
    \(p\) & \(\Per\) & initial & \(b_1/b_0\) & \(b_2/b_0\) & \(b_3/b_0\) & \(b_4/b_0\) \\
    \midrule
    \(1000\) & \(10^3\) & uniform & \(0.0053\) & \(0.0018\) & \(0.0015\) & \(0.0009\) \\
    \(1000\) & \(10^3\) & centre  & \(1.578\)  & \(0.735\)  & \(0.177\)  & \(0.030\)  \\
    \(1000\) & \(10^3\) & wall    & \(-0.663\) & \(0.252\)  & \(-0.057\) & \(0.0087\) \\
    \(100\)  & \(10^1\) & uniform & \(0.0189\) & \(0.0033\) & \(0.0026\) & \(0.0005\) \\
    \(100\)  & \(10^1\) & centre  & \(1.597\)  & \(0.770\)  & \(0.199\)  & \(0.037\)  \\
    \(100\)  & \(10^1\) & wall    & \(-0.656\) & \(0.242\)  & \(-0.051\) & \(0.0068\) \\
    \bottomrule
  \end{tabular}
\end{table}

\begin{figure}
  \centering
  \includegraphics[width=0.98\linewidth]{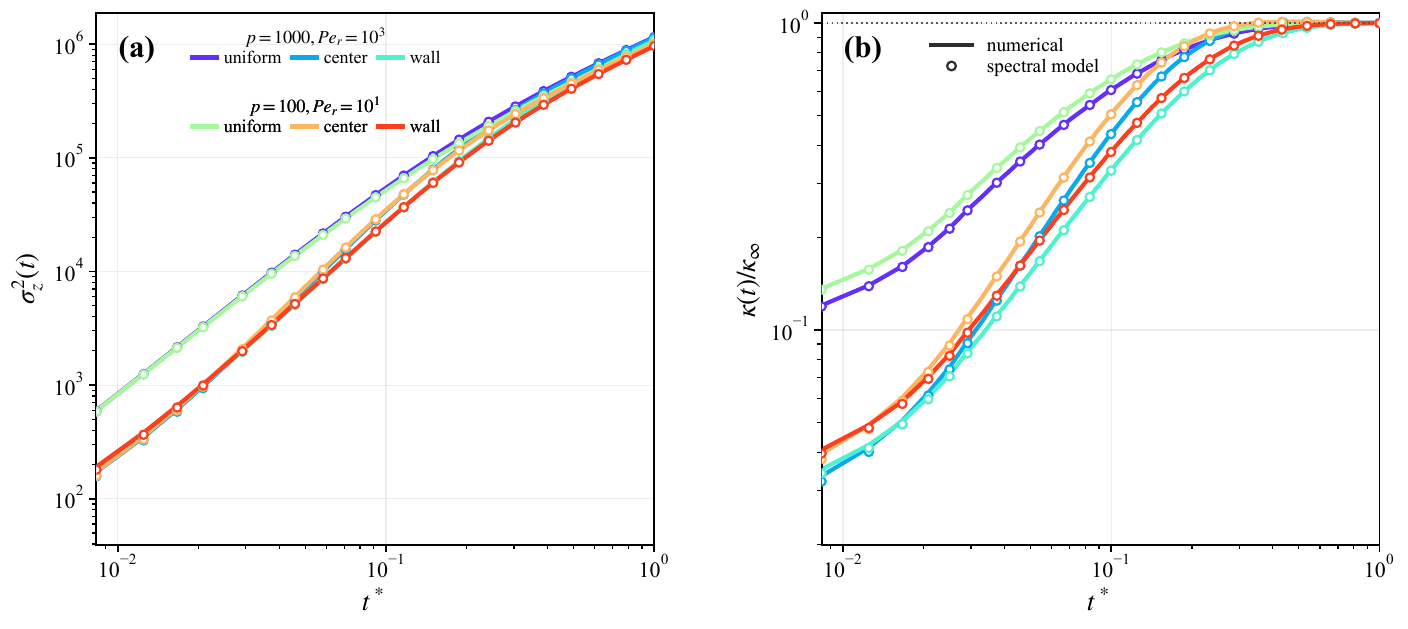}
  \caption{Comparison between the full-time spectral model and the full axisymmetric tensorial equation. Panel (a) compares the axial variance \(\sigma_z^2(t)\). Panel (b) compares the time-dependent dispersion coefficient \(\kappa(t)/\kappa_\infty\), where \(\kappa_\infty\) is the long-time spectral limit for the corresponding parameter set. Solid lines are direct simulations of the full equation, and open circles are spectral-moment predictions from \eqref{eq:moment-system}. Colours denote the three initial radial distributions in Figure~\ref{fig:spectral-initial}; the two parameter sets are \(p=1000,\Per=10^3\) and \(p=100,\Per=10^1\), both with \(\Pe=10^4\).}
  \label{fig:spectral-validation}
\end{figure}

The spectral moment system accurately tracks the full-time variance trajectory across the six validation cases (Figure~\ref{fig:spectral-validation}(a)). The initial radial weight controls the early slope because the centre- and wall-enriched packets sample different parts of the Poiseuille profile before the non-zero modes decay. Once radial relaxation removes that memory, the cases approach a common Taylor--Aris linear-growth regime. The variance slope is reported through the time-dependent dispersion coefficient
\begin{equation}
  \kappa(t)=
  \frac{1}{2\Pe^2}
  \frac{\dd\sigma_z^2}{\dd t}.
\label{eq:kappa-run}
\end{equation}
Across all cases, \(\kappa(t)\) rises from pre-asymptotic values below the long-time limit towards unity (Figure~\ref{fig:spectral-validation}(b)). This evolution records the gradual formation of velocity-deviation correlations under radial sampling. The centre-enriched initial condition mainly samples the fast region, and the near-wall-enriched initial condition mainly samples the slow region, producing different early variance growth. After radial relaxation over the scale \(\lambda_1^{-1}\), the non-zero modes decay and all cases converge to the same \(\kappa_\infty\). Thus \(\kappa(t)\) is a finite-time diagnostic of the evolving velocity correlation, with limit \(\kappa_\infty\).

\begin{table}
  \centering
  \caption{Long-time fitted comparison between the full equation and the spectral moment system in Figure~\ref{fig:spectral-validation}. Here \(\kappa_{\rm full}\) is fitted from the full tensorial equation and \(\kappa_{\rm spec}\) is predicted by the spectral moment system.}
  \label{tab:fig9-validation}
  \begin{tabular}{cccrrr}
    \toprule
    \(p\) & \(\Per\) & initial & \(\kappa_{\rm full}\) & \(\kappa_{\rm spec}\) & relative difference \\
    \midrule
    \(1000\) & \(10^3\) & uniform & \(6.2880718\times10^{-3}\) & \(6.2828222\times10^{-3}\) & \(-0.0835\%\) \\
    \(1000\) & \(10^3\) & center  & \(6.2955961\times10^{-3}\) & \(6.2903464\times10^{-3}\) & \(-0.0834\%\) \\
    \(1000\) & \(10^3\) & wall    & \(6.2813485\times10^{-3}\) & \(6.2760999\times10^{-3}\) & \(-0.0836\%\) \\
    \(100\)  & \(10^1\) & uniform & \(5.4812592\times10^{-3}\) & \(5.4776559\times10^{-3}\) & \(-0.0657\%\) \\
    \(100\)  & \(10^1\) & center  & \(5.4838850\times10^{-3}\) & \(5.4802900\times10^{-3}\) & \(-0.0656\%\) \\
    \(100\)  & \(10^1\) & wall    & \(5.4790428\times10^{-3}\) & \(5.4754328\times10^{-3}\) & \(-0.0659\%\) \\
    \bottomrule
  \end{tabular}
\end{table}

Across the six cases, the maximum relative difference in the long-time dispersion coefficient remains below \(0.084\%\), the maximum absolute difference in fitted mean speed is less than \(6.9\times10^{-5}\), and the final relative mass drift is of order \(10^{-13}\). Thus the radial spectral model governed by \(D_{rr}\) recovers the long-time Taylor coefficient and describes the transition from far-from-equilibrium radial injection to the final linear-growth regime.

\subsection{Extension to power-law non-Newtonian background flow}

The preceding validation follows the approach to Taylor dispersion in Poiseuille flow. The spectral construction depends on two prescribed inputs: the radial mixing operator set by local orientation and the velocity matrix set by axial shear. A useful extension changes the background shear profile while keeping the same Jeffery--Brownian closure. We therefore consider fully developed pipe flow of a power-law generalized Newtonian fluid, a velocity family used to probe transient Taylor--Aris response in tubes \citep{vedel2014}, normalized by the centreline speed as
\begin{equation}
  u_n(r)=1-r^{1+1/n}.
\label{eq:powerlaw-velocity}
\end{equation}
Here \(n\) is the power-law index, with \(n=1\) recovering the Poiseuille profile, \(n<1\) representing shear-thinning behaviour, and \(n>1\) representing shear-thickening behaviour. The local orientation closure is still obtained from the Jeffery--Brownian equation; only the mapping from radius to local shear parameter changes:
\begin{equation}
  q_n(r)=\frac12\Per |u_n'(r)|
  =
  \frac12\Per\left(1+\frac1n\right)r^{1/n}.
\label{eq:powerlaw-q}
\end{equation}
This gives
\begin{equation}
  D(r)=D_{rr}\!\left(q_n(r);p\right),
  \qquad
  A(r)=D_{rz}\!\left(q_n(r);p\right),
  \qquad
  B(r)=D_{zz}\!\left(q_n(r);p\right).
\end{equation}

With these replacements, the reduced-order spectral propagation form in \eqref{eq:alltime-model}--\eqref{eq:moment-system} remains unchanged. The eigenproblem is determined by \(D(r)\), and \(u(r)\) in the velocity matrix is replaced by \(u_n(r)\); the initial projection and moment equations use the same Sturm--Liouville modes. The essential inputs to the spectral model are therefore the radial mixing operator and the velocity matrix. The quadratic Poiseuille profile used in the preceding sections is one member of this axisymmetric class.

The change in background velocity profile can be separated into its effect on initial velocity sampling, radial relaxation and modal coupling (Table~\ref{tab:powerlaw-modal-diagnostics}). For an initial radial weight \(h\), define
\(\bar u_h^{(0)}=\int_0^1 r h(r)u_n(r)\,\dd r/\int_0^1 r h(r)\,\dd r\), and let
\(\Delta u_0=\bar u_{\rm centre}^{(0)}-\bar u_{\rm wall}^{(0)}\). The first-mode Taylor contribution is
\(\kappa_1=U_{01}U_{10}/\lambda_1\). The shear-thinning profile \(n=0.5\) gives a larger zero-mode sampling speed \(U_{00}\), because a larger part of the cross-section remains close to the centreline speed. The shear-thickening profile \(n=1.5\) gives a slightly larger centre--wall initial velocity contrast and concentrates the Taylor response more strongly in the first non-zero mode. The signed first-mode projections of centre and wall injections remain comparable across \(n\), with opposite signs, so the two injections excite the same leading relaxation mode from opposite sides of the radial profile. The background flow therefore changes the transient route mainly through the velocity matrix and the shear-dependent decay rates.

\begin{table}
  \centering
  \caption{Spectral diagnostics for the power-law background-flow cases. Here \(b_{1,c}/b_0\) and \(b_{1,w}/b_0\) are the first-mode projections for centre- and wall-enriched injections, and \(\kappa_1=U_{01}U_{10}/\lambda_1\).}
  \label{tab:powerlaw-modal-diagnostics}
  \begin{tabular}{cccccccc}
    \toprule
    \((p,\Per)\) & \(n\) & \(U_{00}\) & \(\Delta u_0\) & \(\lambda_1\) & \(b_{1,c}/b_0\) & \(b_{1,w}/b_0\) & \(100\kappa_1/\kappa_\infty\) \\
    \midrule
    \((1000,10^3)\) & \(0.5\) & \(0.597\) & \(0.553\) & \(12.31\) & \(1.589\) & \(-0.660\) & \(93.9\) \\
    \((1000,10^3)\) & \(1.5\) & \(0.454\) & \(0.576\) & \(12.12\) & \(1.575\) & \(-0.664\) & \(98.1\) \\
    \((100,10^1)\)  & \(0.5\) & \(0.592\) & \(0.553\) & \(14.04\) & \(1.614\) & \(-0.647\) & \(93.9\) \\
    \((100,10^1)\)  & \(1.5\) & \(0.451\) & \(0.576\) & \(13.85\) & \(1.589\) & \(-0.659\) & \(98.1\) \\
    \bottomrule
  \end{tabular}
\end{table}

\begin{figure}
  \centering
  \includegraphics[width=0.98\linewidth]{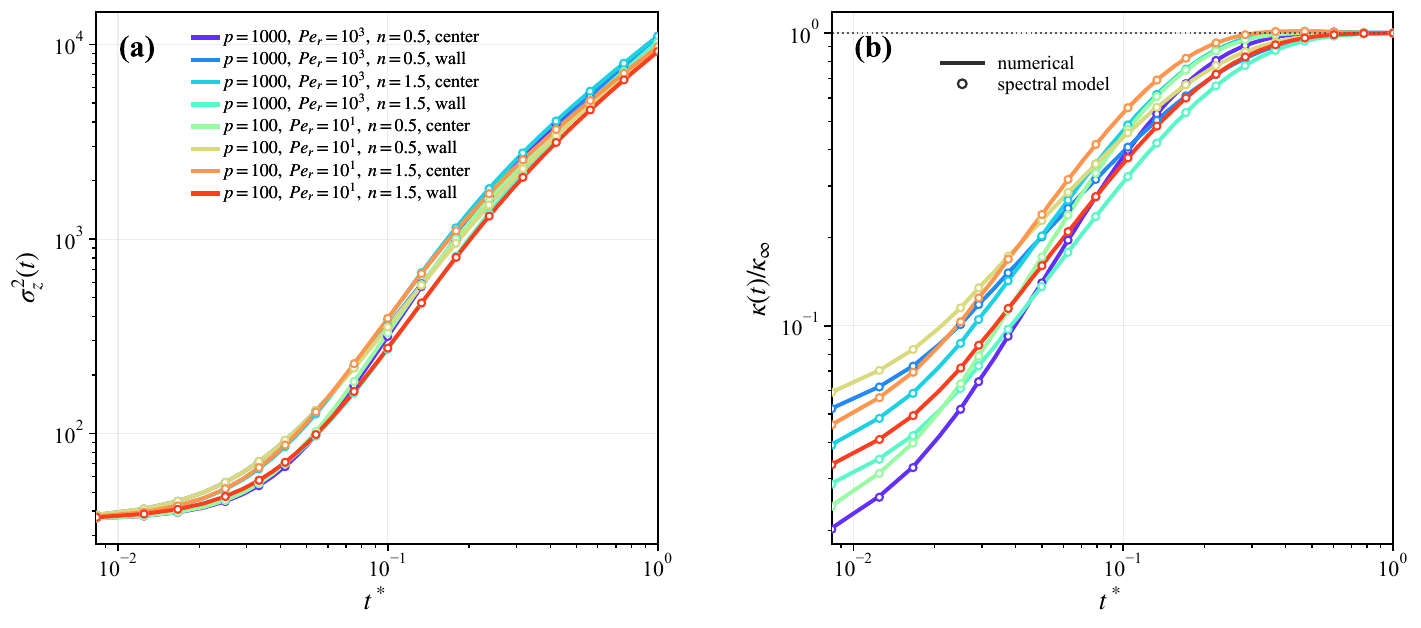}
  \caption{Comparison between the reduced-order spectral model and the full axisymmetric tensorial equation in power-law non-Newtonian background flow. Panel (a) compares the axial variance \(\sigma_z^2(t)\). Panel (b) compares the normalized time-dependent dispersion coefficient \(\kappa(t)/\kappa_\infty\), where \(\kappa_\infty\) is the long-time spectral limit for the corresponding parameter set. Solid lines are direct simulations of the full equation, and open circles are spectral-moment predictions from \eqref{eq:moment-system}. The eight cases include \(p=1000,\Per=10^3\) and \(p=100,\Per=10^1\), power-law indices \(n=0.5,1.5\), and centre-enriched and near-wall-enriched initial conditions, all at \(\Pe=10^4\).}
  \label{fig:powerlaw-nonnewtonian}
\end{figure}

For both shear-thinning and shear-thickening profiles, the spectral moment system follows the variance growth obtained from the full equation (Figure~\ref{fig:powerlaw-nonnewtonian}(a)). Centre-enriched and near-wall-enriched initial conditions sample different velocity regions at early times, leading to different pre-asymptotic slopes. After the radial non-equilibrium modes decay, the curves for a given parameter set enter a common Taylor linear-growth regime. Changing \(n\) modifies the velocity deviation and the distribution of \(q_n(r)\), thereby changing \(D_{rr}(r)\) and the spectral decay rates. These effects enter the moment system through the eigenvalues \(\lambda_m\) and the velocity matrix \(U_{mn}\).

The normalized time-dependent dispersion coefficient gives the corresponding finite-time view (Figure~\ref{fig:powerlaw-nonnewtonian}(b)). In all eight cases, \(\kappa(t)/\kappa_\infty\) approaches unity from pre-asymptotic values controlled by the initial radial memory. The maximum relative difference between the long-time fitted \(\kappa\) and the spectral prediction remains below \(0.12\%\), and mass drift remains at the \(10^{-12}\) level. These results show that, once a steady axisymmetric power-law background shear profile is specified, the same orientation-closure and radial spectral propagation framework applies after replacing \(u(r)\) and \(q(r)\). Cases with particle feedback, concentration-dependent viscosity or non-axisymmetric disturbances require a reformulated background flow and transport closure.

\section{Conclusions}

By connecting local Jeffery--Brownian orientation statistics with a long-wave Taylor--Aris reduction, we have obtained a tensorial description of dilute Brownian rods in circular-tube Poiseuille flow. The local shear \(q=\Per r\) sets the orientation distribution and hence the radius-dependent transport coefficients \(D_{rr}(r)\), \(D_{\phi\phi}(r)\), \(D_{zz}(r)\) and \(D_{rz}(r)\). Placing these coefficients in a conservative axisymmetric transport equation gives the mean migration speed, the cross-diffusive speed correction, the direct axial diffusivity and the leading \(\Pe^2\)-scaled Taylor coefficient from a single asymptotic framework.

The central consequence of the orientation closure is a change in transverse sampling. Streamwise alignment reduces \(D_{rr}\) in high-shear layers, so the leading radial equilibrium is weighted by \(\pi(r)\propto rD_{rr}^{-1}(r)\). The particle cloud therefore samples the Poiseuille velocity through an orientation-controlled measure. This reweighted sampling shifts the velocity-deviation source and strengthens the radial cell response, producing a small non-monotone change in the leading mean speed and a larger increase in the Taylor coefficient. A scalar effective diffusivity would obscure this radial mechanism.

For slender rods, the effect becomes appreciable at large rotational P{\'e}clet number. At \(p=1000\) and \(\Per=10^4\), the leading Taylor coefficient reaches \(\kappa/\kappa_s=1.230\), corresponding to a \(23\%\) enhancement over the classical spherical value. At the same \(\Per\), the \(p\to\infty\) result reaches \(\kappa/\kappa_s=1.304\), close to the fully aligned reference value \(4/3\). The radial decomposition places the excess dispersion mainly in a broad middle-to-outer annulus, approximately \(0.5<r<0.8\), where velocity contrast and the cell-function slope are both substantial.

The tensorial structure also separates the asymptotic roles of the diffusion components. The radial coefficient \(D_{rr}\) controls the invariant measure, radial mixing operator and leading Taylor coefficient. The cross coefficient \(D_{rz}\) enters through conservative radial--axial fluxes and produces the \(O(\Pe^{-1})\) correction to the mean speed. The axial coefficient \(D_{zz}\) contributes the direct diffusivity \(K_{\rm dir}\) to the unscaled variance. Direct simulations of the full tensorial equation support this ordering, giving long-time coefficients and field structures consistent with the reduced theory.

Because the long-time cell problem and finite-time radial relaxation share the same radial mixing operator, the Taylor limit and the decay of radial memory can be treated within one spectral representation. The resulting Sturm--Liouville model captures the variance growth of centre-enriched, cross-section-filling and near-wall-enriched packets before they converge to the common Taylor regime. The same construction applies to steady axisymmetric power-law background flows after replacing \(u(r)\) and the shear mapping \(q(r)\), showing that the reduction is organized around the orientation closure and radial operator, with the quadratic Poiseuille profile appearing as one member of this axisymmetric class.

The reduction assumes dilute passive rods, rapid local orientational relaxation and an imposed axisymmetric background flow. Near-wall hydrodynamic interactions, finite-size exclusion, concentration-dependent stresses, particle feedback and non-axisymmetric disturbances would alter either the local closure or the cross-sectional transport problem. Within these assumptions, the present formulation provides a passive tensorial benchmark for future comparison with particle-scale Brownian-dynamics simulations and controlled microfluidic measurements.

\clearpage
\appendix
\numberwithin{equation}{section}
\renewcommand{\theHsection}{appendix.\Alph{section}}
\renewcommand{\theHequation}{appendix.\Alph{section}.\arabic{equation}}
\renewcommand{\thefigure}{S\arabic{figure}}
\renewcommand{\theHfigure}{appendix.S\arabic{figure}}
\setcounter{figure}{0}

\section[Appendix A. Local orientation closure and semianalytic quadrature]{Local orientation closure and semianalytic quadrature}
\label{app:orientation-solver}

This appendix gives the numerical construction of the local closure and the associated semianalytic radial quadratures. The steady orientation equation \eqref{eq:orientation-fp} is first written using
\begin{equation}
  \mu=\sin\psi,\qquad \dd\Omega=\dd\theta\,\dd\mu,
  \qquad 0\le\theta<2\pi,\quad -1\le\mu\le1 ,
\end{equation}
which gives a tensor-product quadrature rule. The head-tail symmetry of the rod and the symmetry of simple shear are built into the retained real spherical-harmonic basis,
\begin{equation}
  \mathcal B_N=
  \{Y_{2\ell,0},\,Y_{2\ell,2m}^{c},\,Y_{2\ell,2m}^{s}
  :1\le\ell\le N,\ 1\le m\le \ell\}.
\label{eq:app-basis}
\end{equation}
Here \(N\) is the angular truncation level, and the superscripts \(c\) and \(s\) denote cosine and sine real spherical harmonics.
The density is represented as the isotropic state plus this finite expansion. Writing the expansion coefficients as \(\bm a(q)\), projection of the drift-diffusion balance gives
\begin{equation}
  \left(L+2qA\right)\bm a(q)
  =
  -2q\,\bm b ,
\label{eq:app-orientation-linear-system}
\end{equation}
where \(L\) is the projected spherical diffusion operator, \(A\) is the projected Jeffery drift operator, and \(\bm b\) is the forcing produced by applying Jeffery drift to the isotropic density. The entries are assembled by evaluating the basis and its \(\theta\)- and \(\mu\)-derivatives at the quadrature nodes. Rotational diffusion is diagonal in the exact harmonic basis; the quadrature mass matrix is monitored as a practical check on angular resolution.

For each \(q\)-point, the linear system \eqref{eq:app-orientation-linear-system} is solved directly. The reconstructed density is renormalized with the same quadrature rule before the moments in \eqref{eq:local-tensor-definitions} are evaluated. The closure is represented by a monotone cubic interpolation table for \(D_{rr}\), \(D_{\phi\phi}\), \(D_{zz}\), \(D_{rz}\), and \(D_{rz}/D_{rr}\). This table is then mapped to the pipe radius using \eqref{eq:radial-mapping}. The radial integrals for \eqref{eq:um0}, \eqref{eq:uA}, and \eqref{eq:kappa} are evaluated by cumulative trapezoidal quadrature on a refined radial grid.

The reported diagnostics come from identities that are independent of the radial problem: angular normalization, the tensor trace in \eqref{eq:single-orientation-diffusion}, positivity of \(D_{rr}\), positivity of \(D_{zz}\), and positive definiteness of the \(r\)-\(z\) diffusion block. These checks are applied before the local table is used in the pipe-scale quadratures.

\section[Appendix B. Conservative discretization of the full tensorial equation]{Conservative discretization of the full tensorial equation}
\label{app:pde-solver}

Field-level validation is performed by discretizing \eqref{eq:closed-pde-conservation}--\eqref{eq:closed-fluxes}. The stored unknown is the axisymmetric conservative variable
\begin{equation}
  Q(r,z,t)=r\,c(r,z,t).
\end{equation}
On a cell-centred radial grid, the finite-volume update has the form
\begin{equation}
  \frac{\dd Q_{ij}}{\dd t}
  =
  -\frac{\mathcal F_{r,i+1/2,j}-\mathcal F_{r,i-1/2,j}}{\Delta r}
  -\frac{\mathcal F_{z,i,j+1/2}-\mathcal F_{z,i,j-1/2}}{\Delta z}.
\label{eq:app-q-conservation}
\end{equation}
The radial face flux is assembled as
\begin{align}
  \mathcal F_{r,i+1/2,j}
  &=
  r_{i+1/2}
  \left[
  u_{D,i+1/2}c^{\rm up}_{i+1/2,j}
  -D_{i+1/2}\frac{c_{i+1,j}-c_{ij}}{\Delta r}
  -A_{i+1/2}\,\delta_z c_{i+1/2,j}
  \right].
\label{eq:app-rflux-discrete}
\end{align}
Here \(u_D=-D'\), the upwind value \(c^{\rm up}\) is chosen from the sign of \(u_D\), \(D_{i+1/2}\) is a face value of \(D_{rr}\), and \(\delta_z c_{i+1/2,j}\) is a centred axial derivative averaged from the two adjacent radial cells. The axis and wall conditions are imposed by setting
\begin{equation}
  \mathcal F_{r,1/2,j}=0,\qquad
  \mathcal F_{r,N_r+1/2,j}=0 .
\end{equation}
Here \(N_r\) is the number of radial cells. The axial flux uses periodic faces. The advective part is upwinded with speed \(\Pe[u(r_i)-u_f]\), where \(u_f\) is the moving-frame speed used in the periodic computation. The \(B Q_z\) contribution is differenced directly in \(Q\), and the \(D_{rz}\) contribution is evaluated by differencing \(A(r)Q/r\) in the radial direction at the two axial cells adjacent to the face. These choices keep all tensorial terms inside the same conservative divergence in \eqref{eq:app-q-conservation}.

Time advancement is a method-of-lines update with the third-order strong-stability-preserving Runge--Kutta scheme. In high-\(\Pe\) runs, the axial advective translation is applied by conservative periodic remapping of each radial row before and after the Runge--Kutta update of the remaining fluxes. The remap uses the local displacement \(\Pe[u(r_i)-u_f]\Delta t\) and periodic interpolation in \(z\).

Mass, mean position and variance are computed from \(Q\). The mean position is tracked on an unwrapped axial coordinate to remove jumps caused by the periodic computational domain. Long-time slopes of the unwrapped mean and variance give the fitted speed and the fitted \(\Pe^2\)-scaled Taylor coefficient used in the comparisons in Figures~\ref{fig:mean-speed-kappa}, \ref{fig:spectral-validation}, and \ref{fig:powerlaw-nonnewtonian}.

\section[Appendix C. Radial spectral discretization for pre-asymptotic propagation]{Radial spectral discretization for pre-asymptotic propagation}
\label{app:spectral-solver}

The pre-asymptotic spectral calculation follows the construction in \eqref{eq:alltime-model}--\eqref{eq:kappa-spectrum}. This appendix gives the discrete assembly. The input profile is \(D(r)=D_{rr}(\Per r)\) from Appendix~\ref{app:orientation-solver}, together with the prescribed velocity profile \(u(r)\).

Linear finite elements are used on \(0\le r\le1\). On an element \(e=[r_i,r_{i+1}]\) with length \(h_e\), midpoint \(r_e\), \(w_e=(r/D)_e\), and \(b_e=(ru/D)_e\), the element matrices are
\begin{align}
  K^e&=\frac{r_e}{h_e}
  \begin{bmatrix}1&-1\\-1&1\end{bmatrix},\\
  M^e&=\frac{w_eh_e}{6}
  \begin{bmatrix}2&1\\1&2\end{bmatrix},\\
  B^e&=\frac{b_eh_e}{6}
  \begin{bmatrix}2&1\\1&2\end{bmatrix}.
\end{align}
After assembly, the generalized eigenproblem \(K\bm v=\lambda M\bm v\) gives the radial modes. The eigenvectors are normalized in the \(M\) inner product, and their signs are chosen consistently using their radial integral. If \(V\) contains the retained normalized eigenvectors as columns, the retained velocity matrix is \(U=V^TBV\).

For field evolution, an arbitrary initial \(c_0(r,z)\) is first interpolated onto the spectral radial grid when needed. Projection onto modes uses trapezoidal weights for \(\int_0^1 r(\cdot)\,\dd r\). The modal amplitudes are transformed in \(z\) by a discrete Fourier transform, each axial wavenumber is advanced with the matrix exponential in \eqref{eq:fourier-propagator}, and the inverse transform reconstructs the modal amplitudes. Division by \(D(r)\) gives the concentration field.

For moment-only predictions, the block system in \eqref{eq:moment-system} is evaluated by a matrix exponential at the requested output times. The calculation also reports the zero-mode speed \(U_{00}\) and the spectral Taylor sum in \eqref{eq:kappa-spectrum}. Orthogonality of the retained modes, agreement of \(U_{00}\) with \eqref{eq:um0}, and agreement of the spectral sum with \eqref{eq:kappa} are the main truncation checks.

\end{document}